\documentclass[fleqn,usenatbib]{mnras}
\usepackage{etoolbox}
\usepackage{newtxtext,newtxmath}

\usepackage[T1]{fontenc}
\usepackage{ae,aecompl}

\usepackage{graphicx}	
\usepackage{amsmath}	
\usepackage{longtable}

\title[Could magnetised OB stars produce all magnetars?]{Testing the fossil field hypothesis: could strongly magnetised OB stars produce all known magnetars?}

\author[E.I. Makarenko et al.]{
Ekaterina I. Makarenko $^{1}$\thanks{E-mail: makarenko@ph1.uni-koeln.de}
Andrei P. Igoshev,$^{2}$
A.F. Kholtygin$^{3,4}$
\\
$^{1}$I.Physikalisches Institut,  Universit{\"a}t zu K{\"o}ln, Z{\"u}lpicher Str.77, K{\"o}ln D-50937, Germany\\
$^{2}$Department of Applied Mathematics, University of Leeds, Leeds LS2 9JT , UK\\
$^{3}$Saint Petersburg State University, Saint Petersburg, 199034, Russia\\
$^4$Institute of Astronomy, Russian Academy of Sciences, Moscow 119017, Russia\\
}

\date{Accepted XXX. Received YYY; in original form ZZZ}

\pubyear{2020}

\begin{document}
\label{firstpage}
\pagerange{\pageref{firstpage}--\pageref{lastpage}}
\maketitle

\begin{abstract}
Stars of spectral types O and B produce neutron stars (NSs) after supernova explosions. Most of NSs are strongly magnetised including normal radio pulsars with $B \propto 10^{12}$~G and magnetars with $B\propto 10^{14}$~G. A fraction of 7-12~per~cent of massive stars are also magnetised with $B\propto 10^3$~G and some are weakly magnetised with $B\propto 1$~G. It was suggested that magnetic fields of NSs could be the fossil remnants of magnetic fields of their progenitors. This work is dedicated to study this hypothesis. First, we gather all modern precise measurements of surface magnetic fields in O, B and A stars. Second, we estimate parameters for log-normal distribution of magnetic fields in B stars and found $\mu_B = 2.83\pm 0.1$ $\log_{10}$~(G), $\sigma_B=0.65\pm 0.09$ for strongly magnetised and $\mu_B = 0.14\pm 0.5$ $\log_{10}$~(G), $\sigma=0.7_{-0.27}^{0.57}$ for weakly magnetised. Third, we assume that the magnetic field of pulsars and magnetars have $2.7$~DEX difference in magnetic fields and magnetars represent 10~per~cent of all young NSs and run population synthesis. We found that it is impossible to simultaneously reproduce pulsars and magnetars populations if the difference in their magnetic fields is 2.7~DEX. Therefore, we conclude that the simple fossil origin of the magnetic field is not viable for NSs.
\end{abstract}




\section{Introduction}

The origin of magnetic fields in massive stars remains enigmatic. The hypothesis that stellar magnetic field can be fossil was firstly 
proposed by~\cite{Cowling1945}, who showed that time scale of ohmic dissipation of the magnetic field in the stars with masses 
$M > 1.5\,M_{\odot}$ exceeds their lifetime and concluded that stellar magnetic fields could be a remnant of the 
magnetic field of protostellar clouds. The idea that magnetic fields could be fossil relics of the fields presenting in the 
interstellar medium was also argued, for example, by~\cite{Moss2003}. 
Numerical modelling by~\cite{Braithwaite-2004} showed that there exist such stable field configurations surviving over all stellar 
lifetime for simple initial magnetic field configuration. \cite{Braithwaite-2006} confirmed this conclusion for more complex initial 
configurations. \cite{DuezBM10, DuezM2010} showed this result analytically. It is also worth mentioning that
during the evolution convective layers appear in the radiative envelope of massive stars which could be a suitable place for magnetic field generation by dynamo mechanism (see e.g. the LIFE project - the Large Impact of magnetic Fields on the Evolution of hot stars \citealt{Oksala17, Martin2018}). Such dynamo fields can be stable for a long time \citep{Shultz18, Oksala2012}. Neither it is not yet clear how exactly such fields evolve in the radiative layers, nor what their influence is through the life of a star from the main sequence till the NS stage, so in this work we will not take into account the interaction of these fields with the fossil field.

Over the past decades, important progress has been achieved in studying stellar magnetism, mainly due to the equipping of 
large telescopes with high-quality spectropolarimeters. Among those are FORS1/2 at VLT \citep{fors1Appenzeller-1998}, 
ESPaDOnS at CFHT \citep{EspadonsDonati-2003}, Narval at TBL \citep{NarvalAuriere-2003}, HARPSpol at ESO 3.6-m 
\citep{Piskunov-2011} and MSS at 6-meter telescope \citep[Northern Caucasus, Russia,][]{MSSPanchuk-2014}.

On the other hand, the modern method for measuring stellar magnetic fields based on the Zeeman effect let us enhance the number of 
massive magnetic stars \citep[e.g.][]{Donati-2009}. To increase the effective signal-to-noise ratio, many spectral lines can be combined 
together
by so-called least-squares deconvolution (LSD) 
method~(\citealt{Donati1997} a separate implementation was conducted by~\citealt{LSDKochukhov-2010}). 

Another technique which uses multiple spectral lines to increase the S/N ratio is given by~\cite{Hubrig-2003}. Improvements to the methods 
were proposed by~\cite{Hubrig-2014}. In order to analyse the presence of weak stellar magnetic fields the so-called multi-line singular value
decomposition (SVD) method for Stokes profile reconstruction developed by~\cite{Carroll-2012} can be also used.
With this new technique, the magnetic field detection limit has dropped to several gauss and even fractions of gauss. 

Most of the new measurements were added by two large projects. The first project is The Magnetism in Massive Stars (MiMeS) project
\citep[see][]{Wade-2016mimes}. The second is the B fields in OB star (BOB) Collaboration \citep{Schoeller-2017}. 
These projects reveal that an extremely low fraction of massive stars is magnetic. Only about 5-7\% of all stars with radiative 
envelopes in the mass range $1.5 - 50 M_\odot$  have large scale mostly dipolar magnetic fields \citep{Fossati-2015,Grunhut-2013}. More recent observations indicate that up to 10-12\% of all massive stars are strongly magnetic \citep{NeinerBrite}.

But what is the situation with the rest of the massive stars? Are they magnetic? Based on recent measurements the weak (1-10\,G) 
magnetic field of A-F star such as Vega (A0 V, \citealt{Lignieres-2009}), $\beta$~UMa (A1 IV,  \citealt{Blazere16}) and others \citep{Neiner-2017, Blazere2020} one can suppose that all these stars have the root-mean square ({\it rms}) magnetic field in the interval [0.1 - 10]~G and may be called weakly-magnetic stars (also known as "ultra-weak field stars" or "ultra-weakly magnetic stars").  
Moreover, there is already a series of works that prove in different ways that magnetic massive stars have bimodal distribution of magnetic field based on observations (\citealt{Auriere_v, Grunhut17}) and based on theoretical or numerical techniques (\citealt{Cantiello2019, Jermyn2020}).


Recent population synthesis combined both normal radio pulsars and magnetars \citep[e.g.][]{popov2010, Gullon-2015}. Thus, \cite{Gullon-2015} noticed that single log-normal distribution for the initial magnetic field poorly describe the population of isolated radio pulsars and magnetars. Therefore, they suggested that the distribution of the initial magnetic field could be truncated (e.g. due to some magnetic field instabilities in proto-NS).  Alternatively, they suggested that there could be two different evolution paths for massive stars depending on whether they are isolated or born in a binary. Up to 70~per cent of massive stars could go through dynamical interaction in a binary \citep{sana_mult}. Some of these interaction could lead to formation of strongly magnetised massive stars via merger as in \citealt{Schneider2019}, tidal synchronisation \citep{popov2006} or via accretion mass transfer \citep{popov2015,popov2016,popov2016b}. \cite{Schneider2019} analysed a different scenario where strong magnetic fields occur as a result of a merger. They performed three-dimensional magnetohydrodynamic simulations for the merger of two massive main sequence stars. They noticed that this process produces strong magnetic fields compatible with 9~kG. These strongly magnetised merger products could be the magnetar progenitors. Stellar mergers occur in $22_{-9}^{+26}$~per~cent of all binaries \citep{renzo2019}, but it is impossible to quantify currently what fraction of mergers end up strongly magnetised.

Therefore, 
some massive stars could be very different from the bulk population because of their past binary interactions or because of
their significant initial magnetisation. Thus, \cite{ferrario2006} suggested that magnetars could originate from strongly magnetised massive stars and radio pulsars are from weakly magnetised ones which we call the fossil field hypothesis hereafter.
The main goal of this article is to check this hypothesis by comparing the magnetic field  distributions both for massive OB stars
and NSs. 


This article is structured as follows. In Section~\ref{s:data} we gather all available measurements of magnetic fields in massive stars and analyse them using the maximum likelihood technique, in Section~\ref{s:ns} we perform a population synthesis for isolated radio pulsars and magnetars, in Section~\ref{s:res} we show our results and compare them with the fossil field origin. Finally, we discuss different properties of magnetised massive stars and magnetars in Section~\ref{s:discussion} and conclude in Section \ref{s:conclusion}.

\section{Magnetic massive stars}
\label{s:data}
\subsection{Data}

\begin{figure}
\center{\includegraphics[width=\columnwidth]{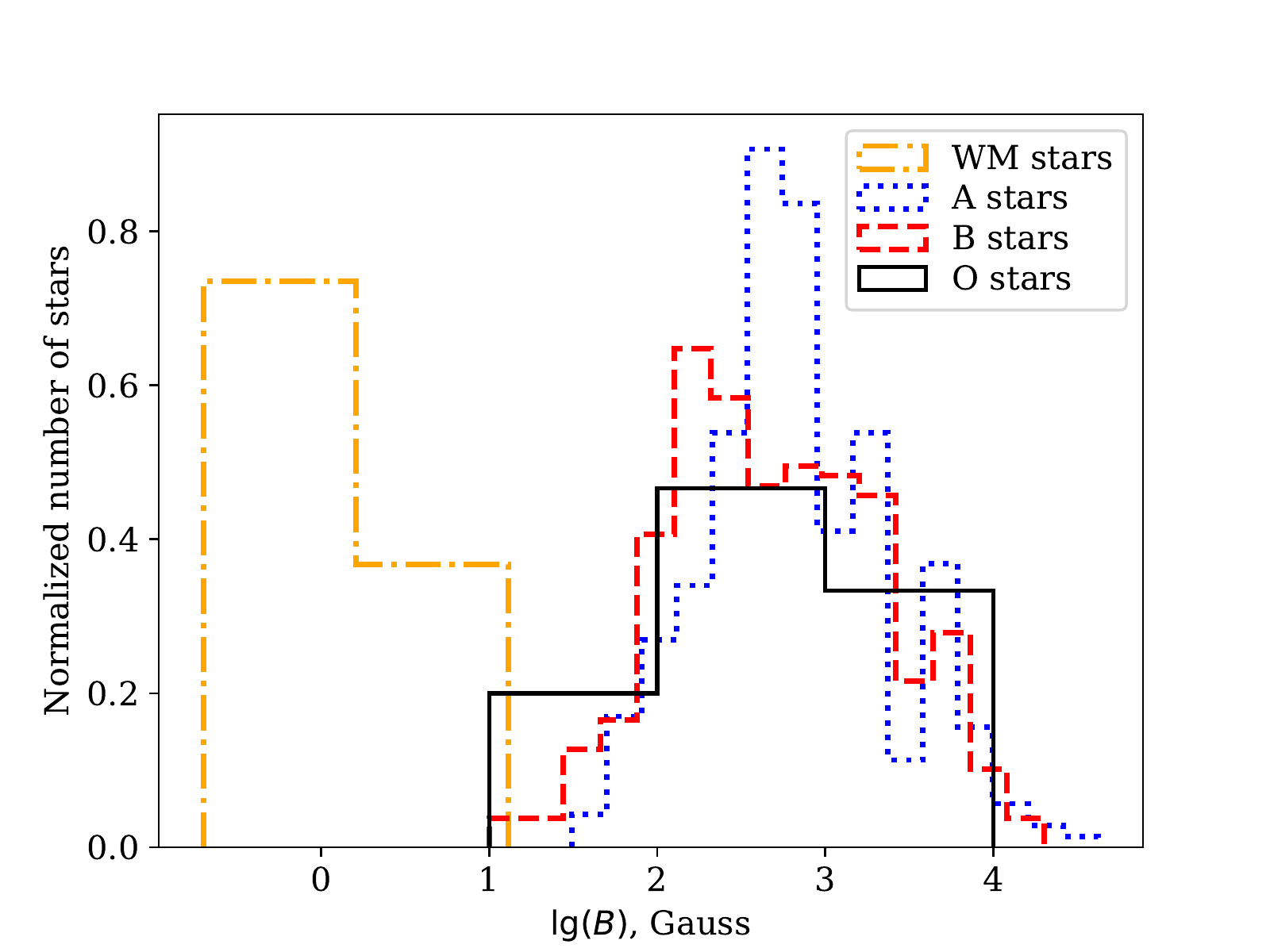}}
\caption{The histogram of magnetic field strengths for O, B, A and weakly magnetic (WM) stars in our sample. The upper limit of the magnetic field of WM stars is 10 G.}
\label{f:distr_OBA}
\end{figure}

Magnetic field measurements were compiled from various sources, see Table~\ref{t:list_of_stars_2} for details. We select only newer measurements (starting from 2006) with a relative error of less than 0.5. 
In this work, we use the value of the root-mean-square (rms) magnetic field \citep{Bohlender1993}:
\begin{equation}
    B = \sqrt{\frac{1}{n} \sum_{k=1}^n (B_l^k)^2},
\end{equation}
where ${B_l}$ is longitudinal magnetic field component, $n$ is number of field measurements. We choose ${B_\mathrm{rms}}$ for this work, because it weakly depends on rotational orientation and 
phase of the star. Based on statistical simulations by \cite{kholtygin2010} it is known that the polar magnetic field ($B_p$) and the rms magnetic field $B_\mathrm{rms}$ are related as $B_\mathrm{rms} \approx 0.2 B_\mathrm{p}$.
Hereafter, ${B}$ means ${B_\mathrm{rms}}$ and the polar magnetic field is labelled with p.
We compile a catalogue which is available in Appendix~\ref{s:catalogue}. We divide stars into strongly and weakly magnetised based on a threshold value of rms magnetic field of 10~G.

We plot the distribution of magnetic fields for different classes of stars in Figure~\ref{f:distr_OBA}. The Kolmogorov-Smirnov test shows that magnetic fields of A and B stars are drawn from two different distributions (
$D= 0.21$,  $p=2 \times 10^{-2}$). At the same time, if we perform the Kolmogorov-Smirnov test for a sample of A stars divided into two equal parts, we obtain approximately the same distribution laws (
$D = 0.19$, $p = 0.28$). Similarly for B stars (
$D = 0.15$, $p =0.63$). It is worth noticing that the KS test by construction does not deal with measurement uncertainties. Therefore, we perform maximum likelihood analysis to estimate parameters of the magnetic field distribution.

\begin{table}
    \centering
    \begin{tabular}{rlc}
        \hline
        \# & Source & Number of stars \\
        \hline
        1&\cite{Alecian14} & 2\\
        2&\cite{alecian16} & 1 \\
        3&\cite{Auriere_v} & 24 \\
        4&\cite{Bagnulo17} & 5\\
        5&\cite{Blazere_alhena} & 1 \\
        6&\cite{Blazere16} & 2 \\
        7&\cite{Castro15} & 1\\
        8&\cite{Cho19} & 30\\
        9&\cite{Davia-Uraz2020} & 1\\
        10&\cite{Folsom2018} & 1\\
        11&\cite{Fossati15} & 1\\
        12&\cite{F2008} & 15\\
        13&\cite{Grunhut13} & 1\\
        14&\cite{Grunhut09} & 1\\
        15&\cite{Hubrig2014} & 3\\
        16&\cite{Hubrig2018} & 1 \\
        17&\cite{Kobzar_v} & 1 \\
        18&\cite{Kochukhov2018} & 1 \\
        19&\cite{Kochukhov19} & 1 \\
        20&\cite{Kurtz2008} & 1\\
        21&\cite{landstreet08} & 8\\
        22&\cite{Lignieres-2009} & 1 \\
        23&\cite{Mathys17} & 17 \\
        24&\cite{Neiner2015} & 1 \\
        25&\cite{Neiner17} & 1 \\
        26&\cite{Petit11} & 1 \\
        27&\cite{Romanyuk18} & 9\\
        28&\cite{Shultz18} & 51\\
        29&\cite{Shultz_17} & 1\\
        30&\cite{Shultz2019} & 1 \\
        31&\cite{Shultz2020_2stars} & 2\\
        32&\cite{Sikora19} & 14 \\
        33&\cite{Sikora16} & 1 \\\
        34&\cite{Silvester12}& 2 \\
        35&\cite{Wade_15} & 1\\
        36&\cite{Wade2012} & 1\\
        37&\cite{Wade_2012} & 1\\
        38&\cite{Wade_11} & 1\\
        39&\cite{Wade_2006} & 1\\
        \hline
        & Total & 208\\
        \hline
    \end{tabular}
    \caption{Literature references for a compiled list of measurements of magnetic fields of massive stars}
    \label{t:list_of_stars_2}
\end{table}

\subsection{Maximum likelihood estimate}
 The maximum likelihood technique must be used to estimate parameters of the magnetic field distribution instead of the simple least-square used in previous works (see e.g. \citealt{Igoshev-2011}) because the latter centres errors on the measured value, while in reality the errors are centred at the actual (unknown) value.
As the initial distribution for magnetic fields, we choose the log-normal distribution with mean $\mu_B$ and standard deviation $\sigma_B$:
\begin{equation}
f(B | \mu_B, \sigma_B) = \frac{1}{B \log 10\sqrt{2\pi}\sigma_B} \exp\left(-\frac{(\log_{10}B - \mu_B)^2}{2\sigma_B^2}\right),  
\label{e:init}
\end{equation}
where $\mathrm{\mu_B}$ is $\mathrm{\log_{10}B_{rms}}$ in Gauss and should be  distinguished from $\mathrm{B_p}$.

The actual magnetic field $B$ is measured as $B'$ because of the measurement process uncertainty. We assume that the distribution of $B'$ as a function of $B$ follows the normal distribution, see e.g. figure 10 by \cite{Hubrig2014}.
Therefore the conditional probability which describes the measurement process is written as:
\begin{equation}
p(B_i' | B) = \frac{1}{\sqrt{2\pi}\sigma_i}\exp\left(-\frac{(B-B_i')^2}{2\sigma_i^2}\right).   
\label{e:cond}
\end{equation}
Here $\sigma_i$ is an error of magnetic field measurement for a particular star and should not be mixed with $\sigma_B$ which is a single unique value describing the distribution for the whole sample.
The joint probability is a multiplication of probabilities eqs. (\ref{e:init}) and (\ref{e:cond}). Because we do not know the value of the actual magnetic field, we integrate it out:
$$
p(B_i'|\mu_B, \sigma_B) = \frac{1}{2\pi\sigma_B\sigma_i\log 10}\int_{B_\mathrm{min}}^{B_\mathrm{max}} \frac{1}{B} \hspace{3.5cm}
$$
\begin{equation}
\hspace{2cm}\times \exp\left(-\frac{(\log_{10}B - \mu_B)^2}{2\sigma_B^2}-\frac{(B-B_i')^2}{2\sigma_i^2}\right)dB.    
\end{equation}
This integral is computed numerically using the grid method with constant logarithmic step because the magnetic field ranges from a fraction of G to tens of kG.
The log-likelihood function is written as:
\begin{equation}
\log \mathcal{L} (\mu_B,\sigma_B) = \sum_{i=1}^N \log (p(B_i'|\mu_B, \sigma_B)).    
\end{equation}
We find the minimum of this function i.e. the maximum of the likelihood. 

\subsection{Result} \label{1}
\begin{figure*}
\begin{minipage}{0.48\linewidth}
	\includegraphics[width=\columnwidth]{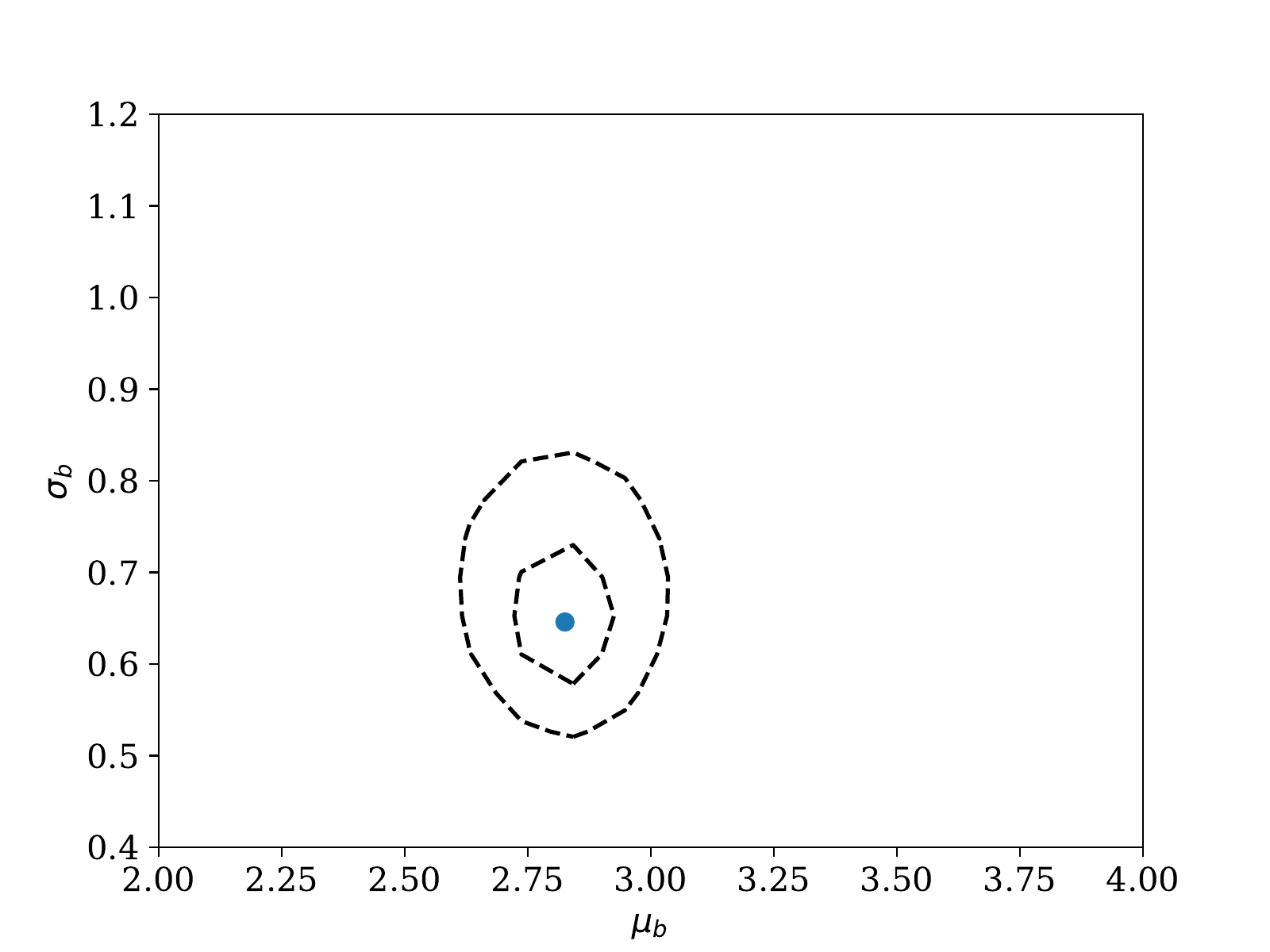}
\end{minipage}
\begin{minipage}{0.48\linewidth}
	\includegraphics[width=\columnwidth]{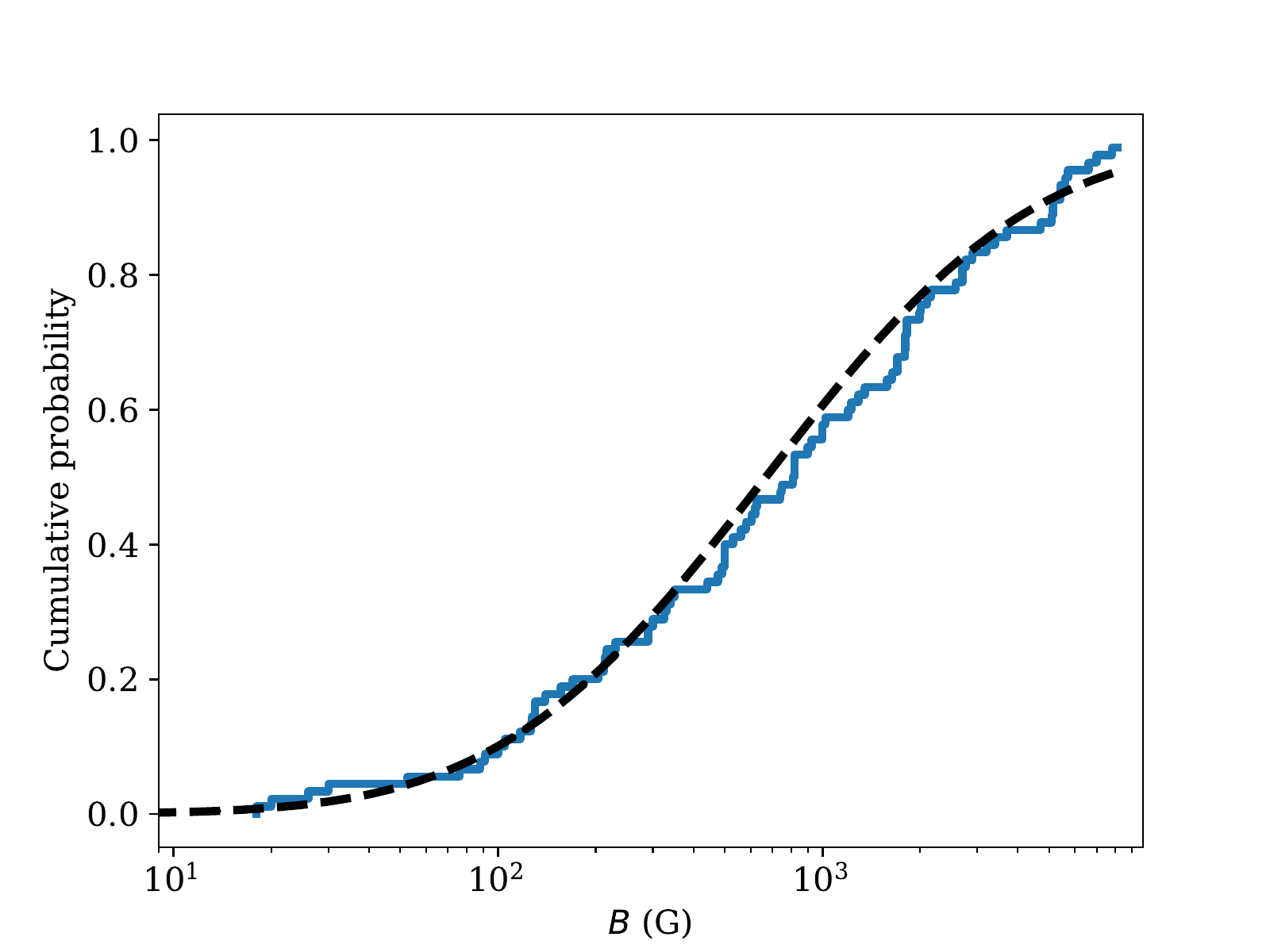}
\end{minipage}
\caption{The position of the maximum likelihood (dot) and credible intervals (70 percent and 99 percent) for parameters of the log-normal distribution for B spectral type stars (left panel). We show the cumulative probability for measured magnetic fields for B stars using a blue solid line and cumulative probability for the best model using a dashed black line. }
    \label{f:B_selected}
\end{figure*}

\begin{figure*}
\begin{minipage}{0.48\linewidth}
	\includegraphics[width=\columnwidth]{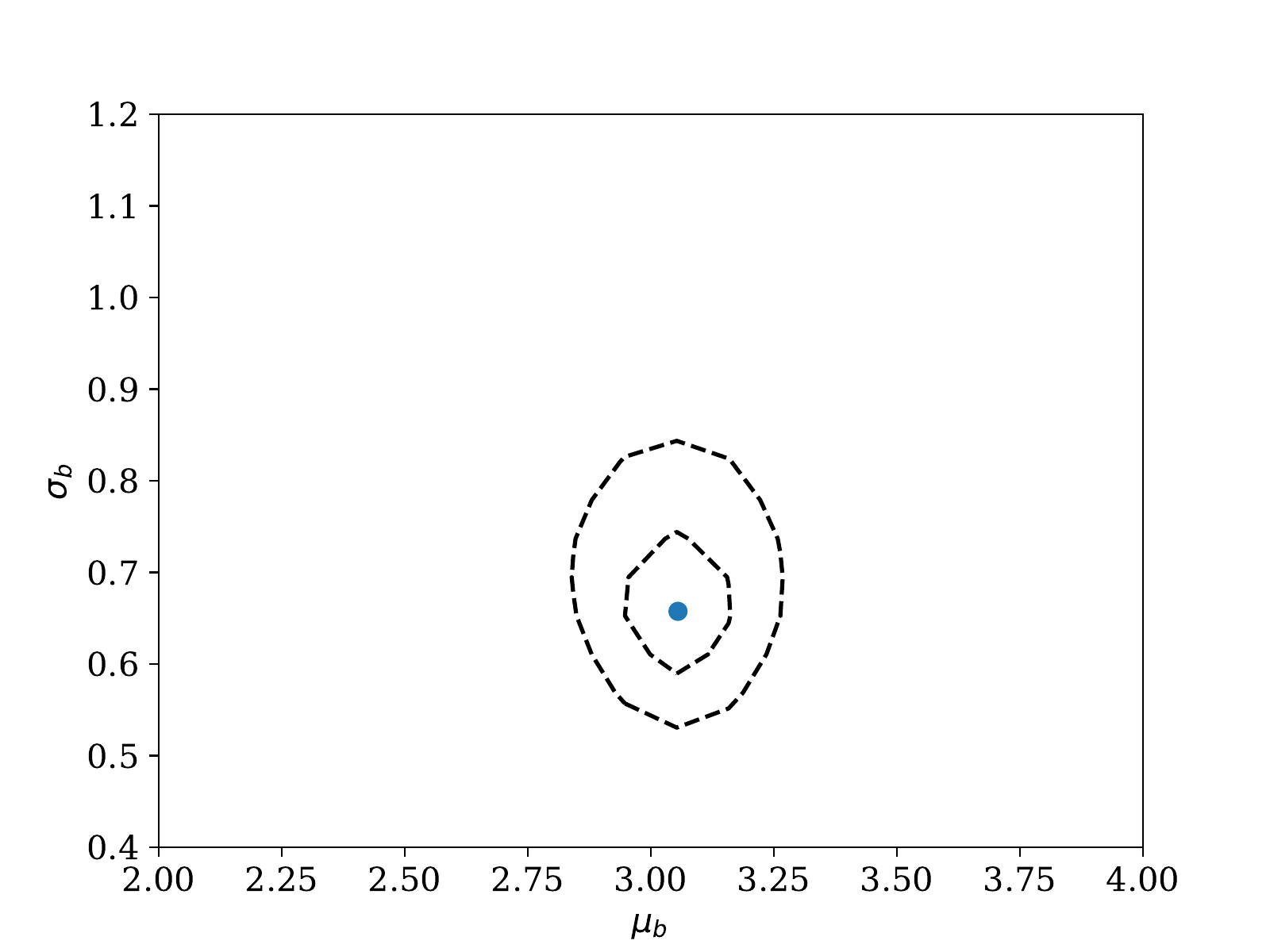}
\end{minipage}
\begin{minipage}{0.48\linewidth}
	\includegraphics[width=\columnwidth]{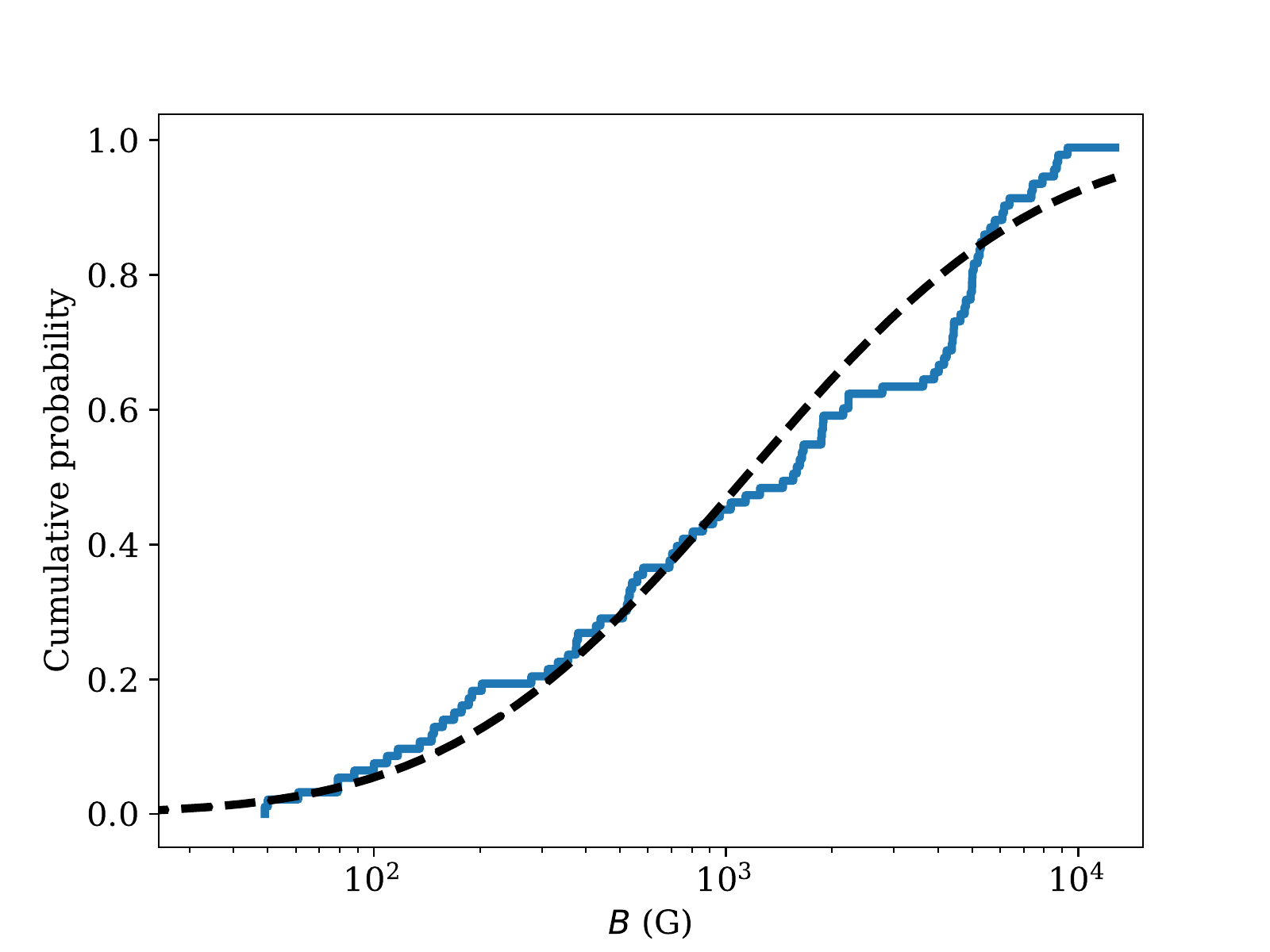}
\end{minipage}
\caption{The same as Fig. \ref{f:B_selected}, but for A stars.}
    \label{f:A_selected}
\end{figure*}

\begin{figure*}
\begin{minipage}{0.48\linewidth}
	\includegraphics[width=\columnwidth]{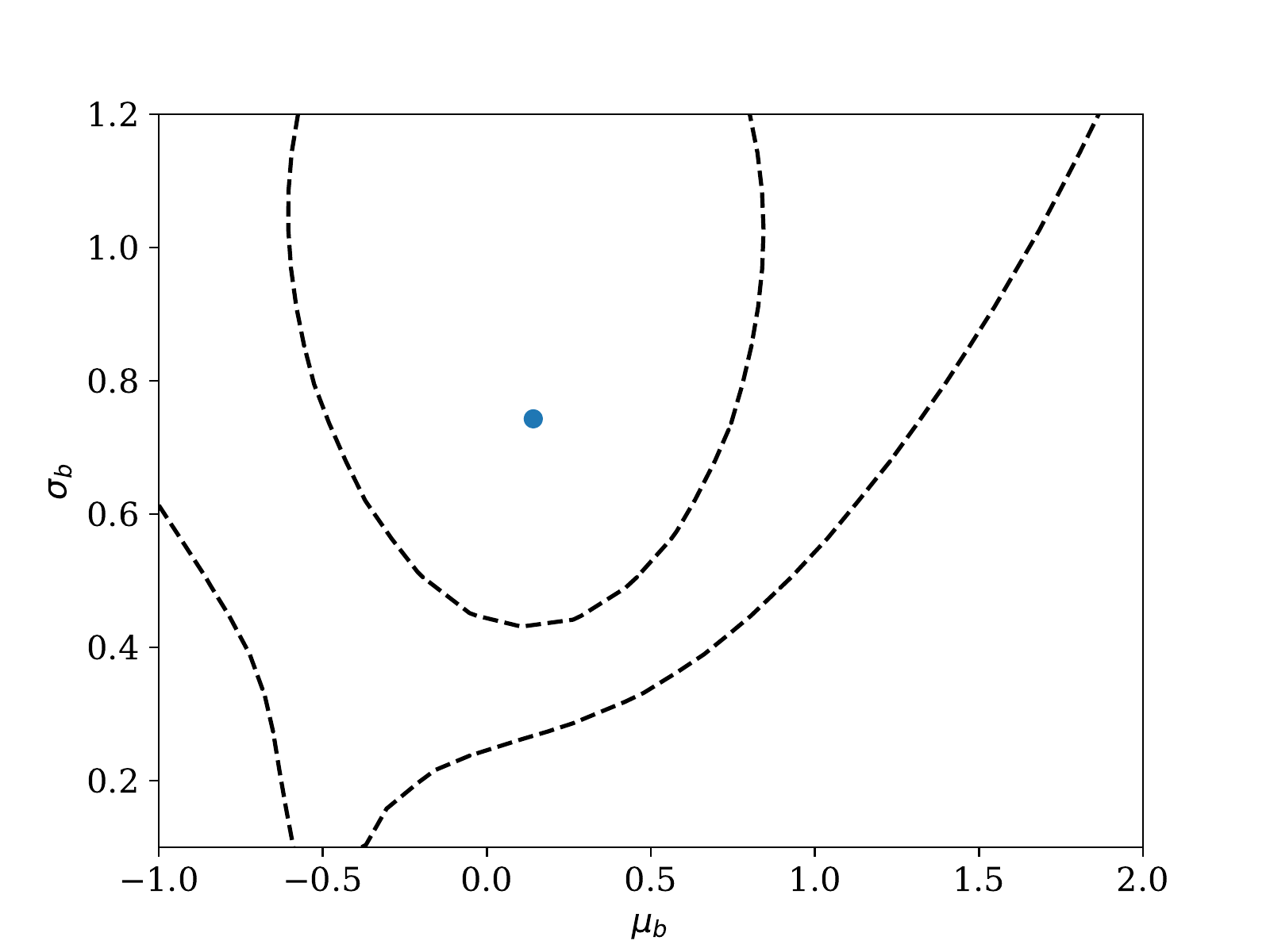}
\end{minipage}
\begin{minipage}{0.48\linewidth}
	\includegraphics[width=\columnwidth]{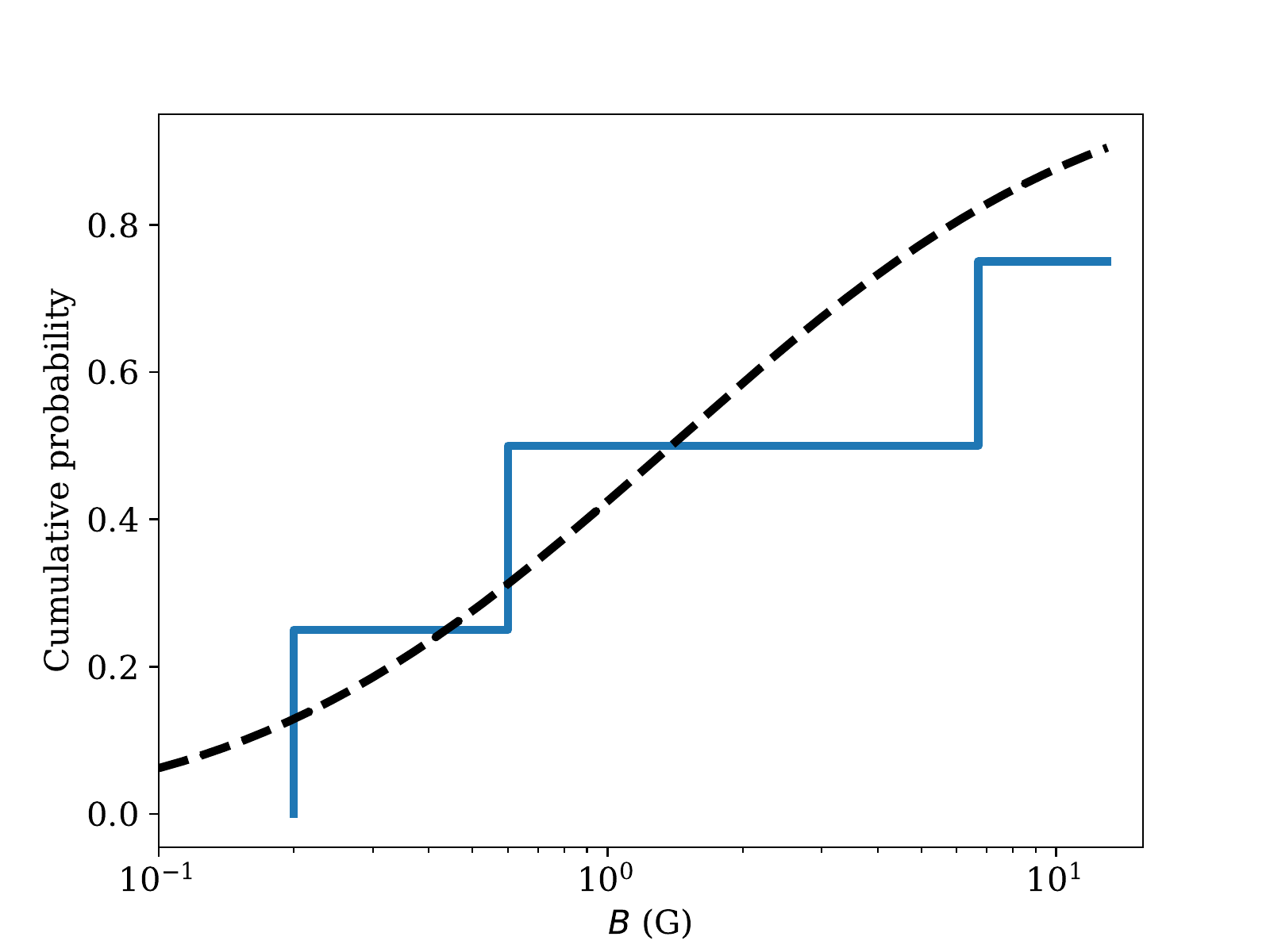}
\end{minipage}
\caption{The same as Fig. \ref{f:B_selected}, but for weakly magnetised stars.}
    \label{f:WM_selected}
\end{figure*}

We summarise the results of the maximum likelihood analysis in Table~\ref{t:max_likelihood} and show the contours of confidence intervals in Figures~\ref{f:B_selected},\ref{f:A_selected} and Figure~\ref{f:WM_selected}. Please note that the number of stars in Table~\ref{t:max_likelihood} does not match the number of stars in the complete catalogue (see Table~\ref{t:list_of_stars_2}), because we selected only those values which satisfy the criterion: >$\mathrm{1.5 \sigma}$. For stars of the spectral type B, the log-normal distribution is a suitable model. After the maximum likelihood optimisation we perform the Kolmogorov-Smirnov test and find the value of $D=0.08$ which corresponds to a p-value of 0.67. Therefore, no significant deviations from log-normality are present.  For A stars the log-normal distribution seems to be marginally consistent with observations: it is worth noticing a large discrepancy between the analytic cumulative distribution and actual distribution of measured magnetic fields in Figure~\ref{f:A_selected} around values of 2-3 kG. The Kolmogorov-Smirnov test which we perform after the optimisation gives $D = 0.14$ which corresponds to a p-value of 0.035. It might be related to a fact that some low-mass A stars might have dynamo \citep{Featherstone2009, TP2021, Seach2020, Zwintz2020}.

In Table~\ref{t:max_likelihood} we analyse weakly magnetised stars separately. The reason to separate them is as follows: stars with a magnetic field of the order of 1 kG constitute 5-10 percent of the massive stellar population. These magnetic fields are easy to measure. Weaker magnetic fields are intrinsically challenging to discover in massive stars due to a limited number of observed absorption lines \citep{Neiner2014}. If weakly magnetised stars were an extended tail of the log-normal distribution seen in massive stars, we would expect magnetic fields of order 1 G to be found in less than $2.82/0.65\approx 4.4\sigma$ cases i.e. in less than 1 in 15000 cases per one strongly magnetised star. In reality, we have 9 measurements (4 good quality for main sequence stars) per 92 strongly magnetised stars and many more are expected to be found with an improvement of instrumentation. Therefore, it is quite safe to assume that a large number (comparable to 50-90 percent of all massive stars) might have weak surface magnetic fields that are hard to measure. We characterise them using an available sample of weakly magnetised stars even though this sample is small and incomplete. We should be cautious and remember that there are stars with non-detected magnetic fields even despite  a very low detection threshold, see \cite{Neiner14_AA}.

\begin{table}
    \centering
    \begin{tabular}{cccccc}
    \hline
    Spectral type & N &$\mu_B$ & $\sigma_B$ & $-2\log \mathcal{L}^*$  \\
                  & & $\log_{10} \mathrm{[G]}$ \\
    \hline
    O             & 10 & $2.62\pm 0.16$ & $0.26^{+0.24}_{-0.11}$        & 137 \\
    B             & 91 & $2.82\pm 0.10$  & $0.65\pm 0.09$        & 1416\\
    A             & 93 & $3.06\pm 0.11$ & $0.66\pm 0.07$        & 3414\\
    Weak magnetic & 4  & $0.14\pm 0.5$ & $0.7_{-0.27}^{+0.57}$ & 15  \\
    \hline
    \end{tabular}
    \caption{Result of the maximum likelihood analysis. N is number of stars of particular spectral type. Uncertainties are 70~per~cent confidence intervals. We selected only measurements which are $>1.5\sigma$.}
    \label{t:max_likelihood}
\end{table}

It is interesting to note that the standard deviation of the log-normal distribution for B stars is $0.65\pm 0.09$ which is within 1-sigma interval of standard deviation independently derived for radio pulsars $\sigma_{\log B} = 0.55$ \citep{2006ApJ...643..332F}. As for the shift of distribution following argument applies: if we assume that $\log_{10} B = 2.62$ is representative of a magnetic field for a typical B2V with the radius of $5.3R_\odot$ and during a collapse, radius shrinks to 10~km, the magnetic field could be amplified many orders of magnitude and reach $\mu_B\approx 14$ which is typical for magnetars. We discuss this possibility in more details in Section~\ref{s:fields_conv}. It is worth to note that magnetic field might decay during the lifetime of massive star. Such possibility was discussed by \cite{Shultz19}. \cite{medvedev2017} tried to estimate the magnetic field decay timescale and found that it exceeds the half of main sequence lift-time. Therefore, we do not expect a very strong decay before NS formation.

\section{Magnetic Neutron stars}
\label{s:ns}

\subsection{Data}
In order to compare the results of our analysis with real pulsars and magnetars we use the ATNF catalogue\footnote{http://www.atnf.csiro.au/research/pulsar/psrcat} \citep{atnf}, McGill Online magnetar catalogue\footnote{http://www.physics.mcgill.ca/~pulsar/magnetar/main.html} \citep{mcgill} and  Magnetar Outburst Online Catalog\footnote{http://magnetars.ice.csic.es/} \citep{zelati}. For detailed comparison of X-ray fluxes we use data by \cite{Mong18}  and \cite{vigano2013} derived for X-ray energy range 2-10~keV and 1-10~keV respectively.

\subsection{Population synthesis of massive stars and NSs}
In this research, we use the population synthesis code NINA\footnote{Source code and documentation are publicly available https://github.com/ignotur/NINA} (Nova Investigii Neutronicorum Astrorum sentence in Latin standing for New Study of Neutron Stars). The simulation algorithm is mostly similar to one presented by \cite{2006ApJ...643..332F} with a few small changes.

We draw masses and positions of massive stars instead of positions of NS. We fix the birthrate at the level of $n_\mathrm{br} = 7$ stars in mass range 8-45~$M_\odot$ per thousand years. Therefore, we draw 7 stars every thousand years adding a random shift to their birth time to spread them uniformly inside the thousand years interval. 
Masses of stars are drawn from the \cite{salpeter1955} initial mass function. We draw positions equally probably from one of four spiral arms, using the same parameters are \cite{2006ApJ...643..332F}. The initial metallicity of a star depends on its birth-location according to \cite{metallicity}. The spiral pattern rotates with speed 26~km~s$^{-1}$~kpc$^{-1}$ \citep{spiral_pattern_rotII, spiral_pattern_rot}. Peculiar velocities of massive stars are drawn from Maxwellian with $\sigma = 15$~km~s$^{-1}$.

Further, we compute lifetime on the main sequence, core mass at the end of evolution and radius of the star using equations derived by \cite{2000MNRAS.315..543H}. The mass and position for NSs are computed based on the parameters of massive stars. 
The natal kick of NSs is drawn from the sum of two Maxwellians distribution according to \cite{2017A&A...608A..57V,igoshev2020} in a form:
\begin{equation}
f(v) = w \sqrt{\frac{2}{\pi}}\frac{v^2}{\sigma_1^3} \exp\left(-\frac{v^2}{2\sigma_1^2}\right) + (1-w)\sqrt{\frac{2}{\pi}}\frac{v^2}{\sigma_2^3} \exp\left(-\frac{v^2}{2\sigma_2^2}\right), 
\end{equation}
with uniform distribution of the natal kick orientation on a sphere. We use parameters $w = 0.42$, $\sigma_1 = 75$~km~s$^{-1}$ and $\sigma_2 = 316$~km~s$^{-1}$.

We integrate the motion of NSs in the Galactic gravitational potential \citep{KuijkenGilmore} using the fourth-order Runge-Kutta method. 
We assume that the magnetic field of normal radio pulsars does not decay based on recent research \citep{2019MNRAS.482.3415I}. For magnetar candidates ($B > 10^{13}$~G and age less than 1 Myr) we post-process result and model magnetic field decay using the same scheme as was described in works by \cite{2018MNRAS.473.3204I,2015AN....336..831I}. This modelling includes the inner crust impurity parameter $Q$ which determines the magnetic field decay timescale due to Ohmic losses. It is believed that impurity parameter is large $Q\gg 1$ \citep{2013NatPh...9..431P} for magnetars since it provides a convenient explanation for the absence of magnetars with a period longer than 12 sec.

We compute radio luminosity of pulsars using the same technique as described by \cite{2006ApJ...643..332F} and we implement exactly the same conditions for the pulsar detectability. The dispersion measure is calculated using the source code by \cite{ymw2017}. For the sky map temperature at radio frequency 1.4 GHz we use map by \cite{dinnat2010map}.

To compare the population of synthetic radio pulsars with observations we bin the $P$~--~$\dot P$ into 20 rectangular areas and compute the C-statistics for each individual bin, see Appendix~\ref{s:statistics} for details. The advantage of C-statistics in comparison to 2D KS test used by \cite{Gullon-2015} is that it is more sensitive to bins with a small number of pulsars. The use of C-statistics is also better justified because the drawing process for pulsars in the particular bin should follow the Poisson distribution. Based on the comparison between $P$~--~$\dot P$ of the Parkes and Swinburne surveys with the modern ATNF data, we conclude that C-statistics difference up to 8-12 means that we can reproduce the sample of observed radio pulsars.

\subsubsection{Magnetic fields of NSs in the case of fossil origin}
\label{s:fields_conv}
\begin{figure*}
	\includegraphics[width=2.0\columnwidth]{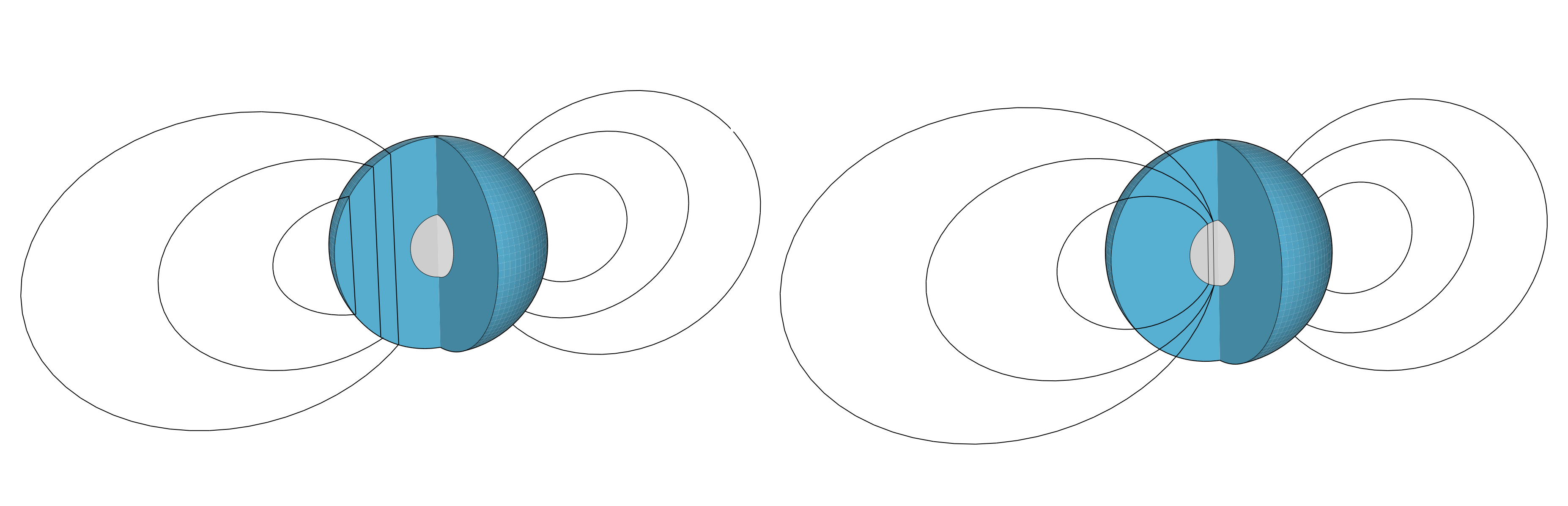}
\caption{Schematic view of magnetic field lines inside and outside of a massive star. The blue region is a radiative envelope, the grey region is a convective core and solid lines show the magnetic field lines.}
    \label{f:mf_lines}
\end{figure*}

The configuration of the magnetic field inside a massive star is complicated. A stable configuration most probably includes both poloidal and toroidal components \citep{Braithwaite-2006}. Here we simplify it significantly to consider only extreme cases. More detailed analysis requires numerically expensive magnetohydrodinamical modelling.
We choose two limiting scenarios: (1) only the core of the star is a good conductor, so currents flow only there (see right panel of the Figure~\ref{f:mf_lines}) and (2) whole star is a uniformly conducting body (left panel of the same figure).

\paragraph{Magnetic flux conservation}
In earlier works, e.g. \cite{Igoshev-2011} researchers considered a conservation of magnetic flux defined as $F = 4\pi B_\mathrm{pole} R_*^2$ where $R_*$ is the stellar radius and $B_\mathrm{pole}$ is the magnetic field at the pole. \cite{kholtygin2010} noticed that the magnetic field at the pole of a massive star is $\alpha \approx 5$ times stronger than the \emph{rms} field measured based on spectral lines. In studies of radio pulsars,  researchers commonly use magnetic field strength at NS equator \citep{handbook} which is two times smaller than the field at the pole for dipolar configuration. Therefore, we use a coefficient $\alpha/2$.

After the supernova explosion, NS has a radius of $R_\mathrm{ns}\approx 10$~km which means that its magnetic field is:
\begin{equation}
B^\mathrm{NS} = \frac{\alpha}{2} B_\mathrm{rms} \left(\frac{R_*}{R_\mathrm{ns}} \right)^2.   
\label{e:flux_conservation}
\end{equation}

Here we consider an example O star with $R_* = 10$~R$_\odot$. If this star is strongly magnetised it can produce NS with mean $B^\mathrm{NS}_p = 5\cdot 10^{14}$~G. If this star is weakly magnetised, it produces an NS with mean $B^\mathrm{NS} = 1.7\cdot 10^{12}$~G. We use these parameters in our model D for $\mu_B$.
On the other hand, if we consider a B star (B2V, R = 5 R$_{\odot}$), we obtain: $B^\mathrm{NS} = 2\cdot 10^{14}$~G and $B^\mathrm{NS} = 4.2\cdot 10^{11}$~G.

\paragraph{Core conductivity}
\label{s:core_conduct}
We assume that the general configuration of magnetic field is a dipole (see Figure~\ref{f:mf_lines}, right panel). In this case, the magnetic field at the pole of a core is stronger than the magnetic field at the pole of a massive star:
\begin{equation}
B_\mathrm{core, pole} = 2 B_\mathrm{p} \left(\frac{R_*}{R_\mathrm{core}}\right)^3.    
\end{equation}
In this case the NS magnetic field at equator is:
\begin{equation}
B^\mathrm{NS} = B_\mathrm{p} \left(\frac{R_*}{R_\mathrm{core}}\right)^3 \left(\frac{R_\mathrm{core}}{R_\mathrm{ns}}\right)^2.   
\label{e:core_cond}
\end{equation}
We take two typical B-stars (B2V, $R_*$ = 5 R$_{\odot}$, $R_\mathrm{core}=0.3R_*$ D. Sz{\'e}csi, 2020, private communication) : (1) strongly magnetic with $B_{\mathrm{star}} \approx$ 700 G and (2) weakly magnetic with $B_{\mathrm{star}}$ $\approx$ 1.4 G and calculate the magnetic field expected at the NS stage as B$^{\mathrm{NS}} \approx 5 \cdot 10^{11}$ G and
B$^{\mathrm{NS}} \approx 2.8 \cdot 10^{14}$ G. These values are nearly identical to ones we obtain for B star in an assumption of magnetic flux conservation. 
These magnetic fields could correspond to normal radio pulsars and magnetars. So, we proceed with this model further and show results of the population synthesis in Section~\ref{s:res}, model E.

\paragraph{Uniformly conducting star}
In this case the magnetic field lines inside the star are straight (see Figure~\ref{f:mf_lines}, left panel).
The magnetic field of the core is a fraction of the surface magnetic field strength which is proportional to the fraction of volume of the core to the volume of the star.
The magnetic field at the core can be computed as:
\begin{equation}
B_\mathrm{core, pole} = 2 B_\mathrm{p}  \left(\frac{R_\mathrm{core}}{R_*}\right)^3.
\end{equation}
In this case the NS magnetic field at equator is:
\begin{equation}
B^\mathrm{NS} = B_\mathrm{p} \left(\frac{R_\mathrm{core}}{R_*}\right)^3 \left(\frac{R_\mathrm{core}}{R_\mathrm{ns}}\right)^2.   
\end{equation}
In reality the radiative envelope is less dense than the core and is consequently less conductive. Additional toroidal magnetic field makes the situation even more complicated.

Again, if we take two typical B-stars and calculate the magnetic field for NSs, we get $B^{\mathrm{NS}}$ $\approx$ $4 \cdot 10^{8}$ G,
$B^{\mathrm{NS}}$ $\approx$ $2 \cdot 10^{11}$ G. We can immediately conclude that this scenario is extremely unrealistic, since the resulting average magnetic fields of NSs are several orders of magnitude lower than the real ones. Therefore, we do not consider this model anymore.

\subsection{Population synthesis of magnetars}

The observational selection for magnetars is very different from one for normal radio pulsars, see e.g. \cite{Gullon-2015}. Magnetars are often discovered during outbursts in X-rays and $\gamma$-rays and confirmed later on by follow up observations aimed at the determination of the rotational period and period derivative. 
Unlike pulsars, magnetars often do not emit radio and follow up observations happen in X-rays to characterise the quiescence state. Therefore, the absorbed X-ray flux in quiescence is the main observational limitation.

The X-ray flux in quiescence depends on magnetic fields which is not constant. The magnetars are believed to lose a significant part of their magnetic energy in $\approx1$~Myr \citep{2013NatPh...9..431P,2018MNRAS.473.3204I}. 

We perform population synthesis as following: first, we introduce a toy model for magnetar population synthesis to illustrate different factors that play an essential role in the population synthesis, and then in Section~\ref{s:realistic_magnetars} we compute a more realistic model.

\subsubsection{Toy models for X-ray luminosity}
\label{s:toy_xray}

We use a toy model to illustrate the importance of X-ray luminosity for the population synthesis of magnetars.
In this toy model we assume no magnetic field decay and choose an initial distribution for magnetic fields in form similar to \cite{2006ApJ...643..332F} i.e. $\mu_B = 12.65$ and $\sigma_B = 0.5$. From the results of pulsar population synthesis we select objects with $B_0 > 10^{13}$~G and age $< 10^6$~years. A similar complete model was computed by \cite{Gullon-2015} as model A.

Observed X-ray flux of magnetars is affected by complicated energy redistribution in the X-ray spectra and absorption in the Galaxy. We keep details of this process until Section~\ref{s:xray_spectra}, but here we consider the dependence of luminosity $L_X$ and flux $S_X$ on the initial magnetic field and age of the magnetar. Because we do not have access to results of detailed magneto-thermal NS evolution with cooling, we use an empirical model which we calibrate using magnetar observations and figure 11 in \cite{vigano2013}. We recommend future researchers who simulate the cooling of magnetars to make the results of their work publicly available to simplify magnetar population synthesis.

\begin{figure}
	\includegraphics[width=\columnwidth]{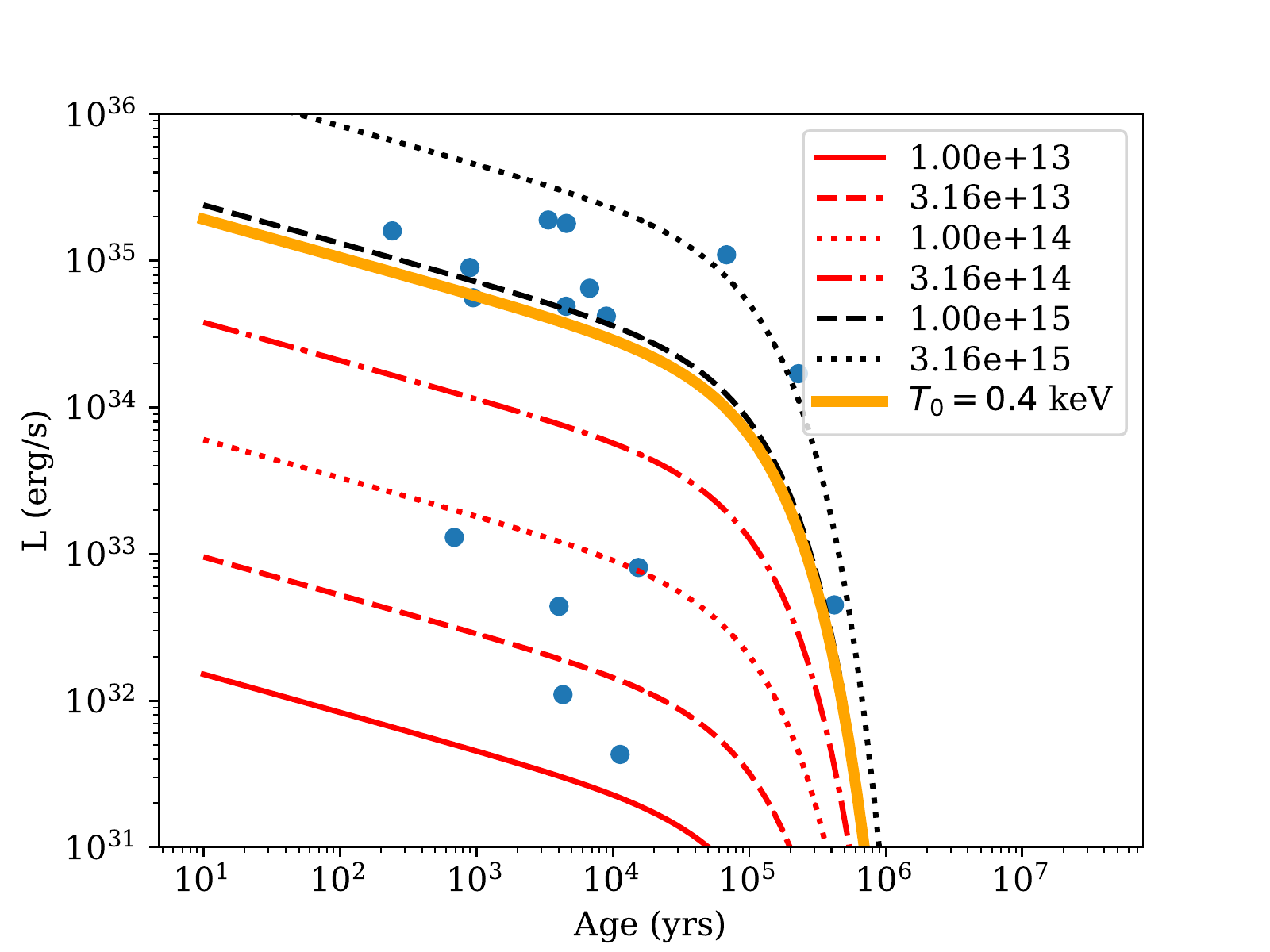}
\caption{X-ray luminosity as a function of time and initial magnetic field in Gauss. The legend shows the values of the initial magnetic field in Gauss. Blue dots show luminosities derived for magnetars in the McGill catalogue.}
    \label{f:lumdec}
\end{figure}

We study evolution of luminosity in the form: $L_X (B_0, t) = L_1 (B_0) L_2 (t)$. According to figure 11 in \cite{vigano2013} the temperature evolves first as power-law of time and start decaying exponentially at timescales of $\propto 10^5$~years. So, in the first toy model we assume the initial temperature $T_0 = 0.4$~keV = $4.6\times 10^6$~K. We further describe the evolution of luminosity as:
\begin{equation}
L (t) = 4\pi R^2 \sigma T_0^4 \left(\frac{t}{1\; \mathrm{year}}\right)^{-0.26} \exp \left(-\frac{t}{10^5\; \mathrm{years}}\right).    
\end{equation}
We plot the evolution of luminosity for this model in Figure~\ref{f:lumdec}. It is immediately obvious that despite quite a large initial temperature, the model misses some magnetars. There is an intrinsic spread in X-ray luminosities seen for magnetars of similar age.

\begin{figure}
\center{\includegraphics[width=\columnwidth]{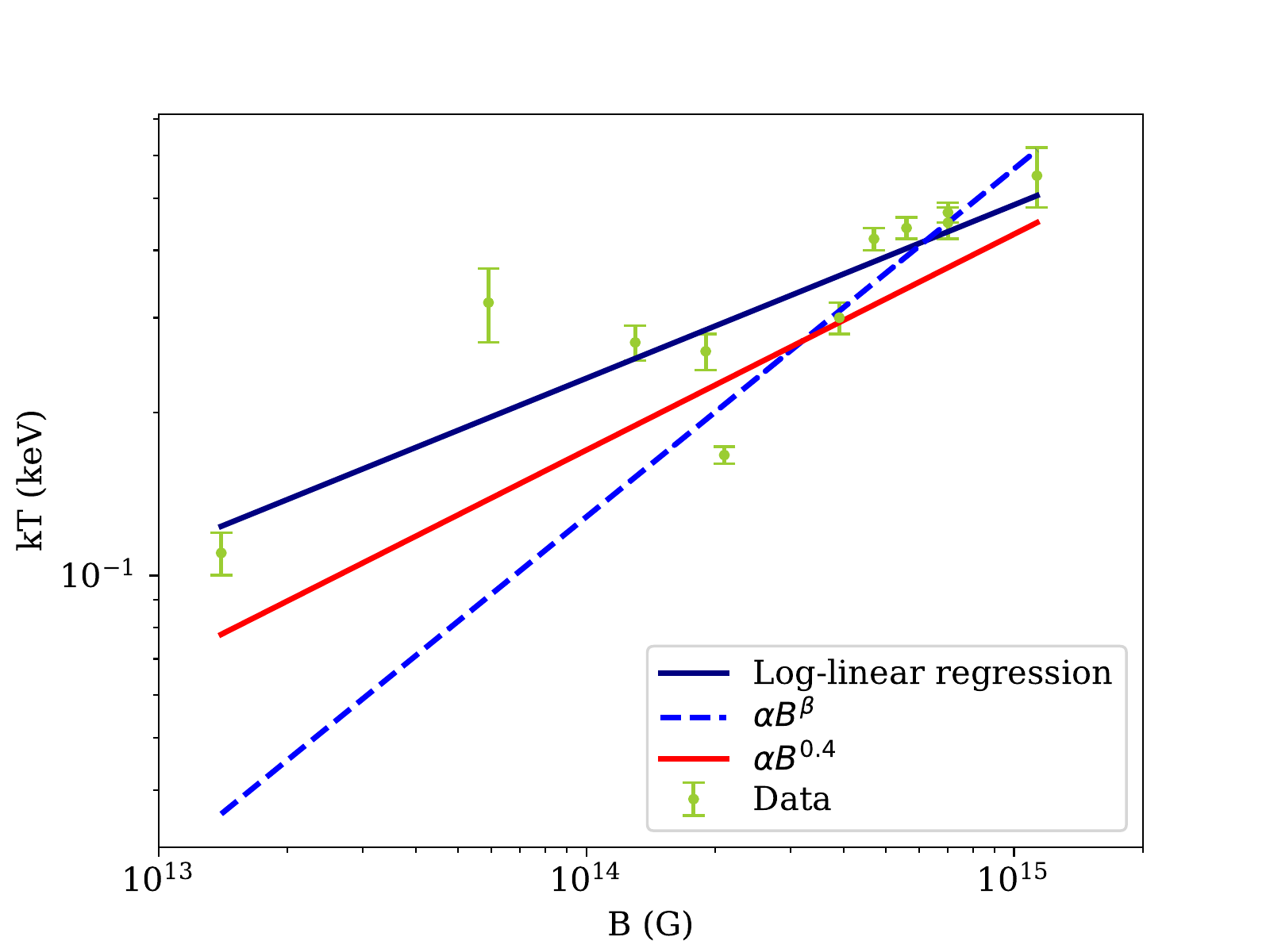}}
\caption{The $kT - B$ dependence for real observed magnetars. Different models of approximation are described in the legend of the graph. The axes are logarithmic.}
\label{f:kTB}
\end{figure}

\begin{figure}
	\includegraphics[width=\columnwidth]{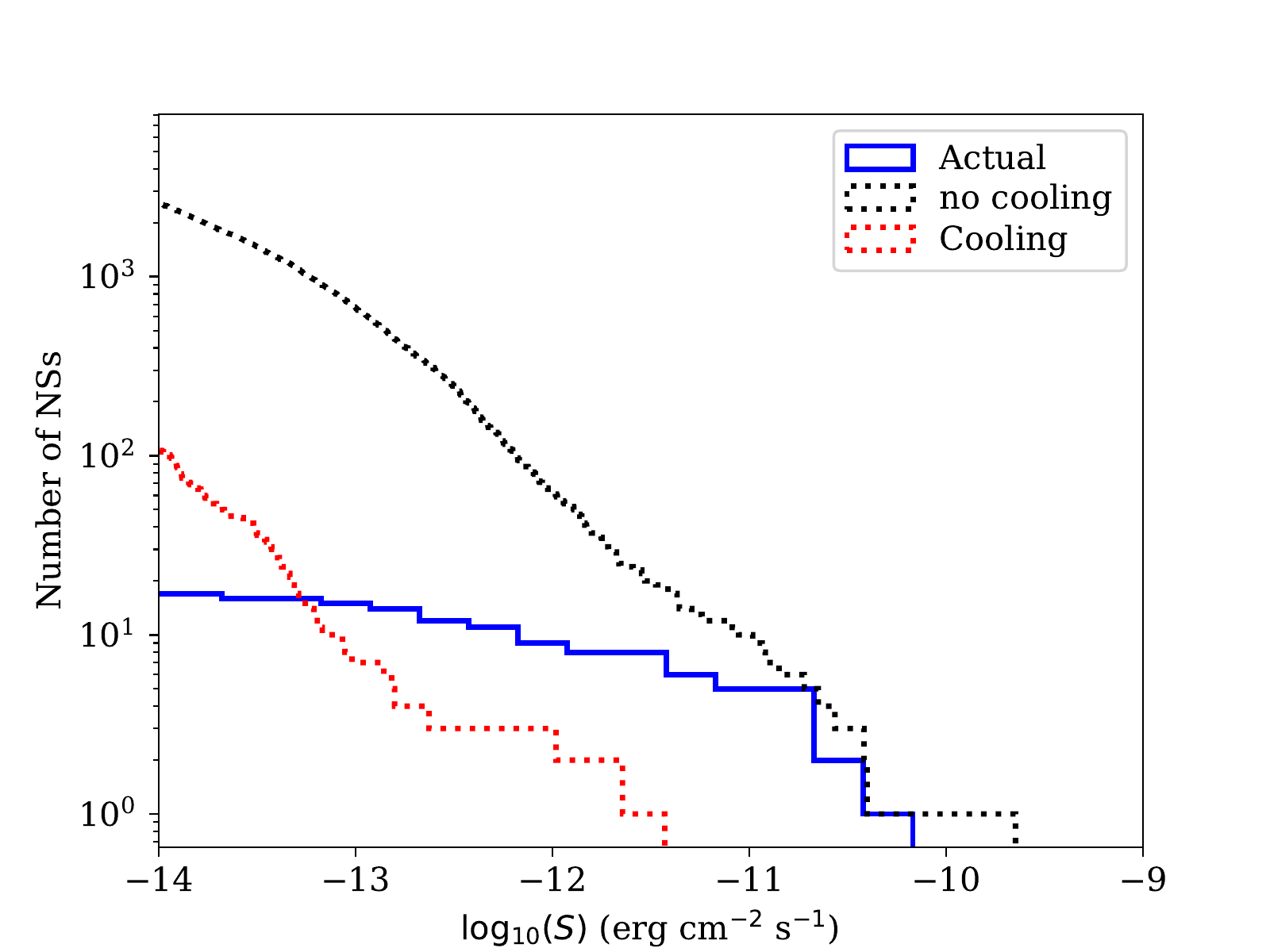}
\caption{X-ray fluxes of magnetars obtained in two toy models. The solid blue line corresponds to the distribution of X-ray fluxes of magnetars in quiescence from the McGill catalogue. THe dotted red line shows X-ray fluxes seen in a model with exponential decay of luminosity with time. The dotted black line shows X-ray fluxes seen in a model where the luminosity decay only due to decay of the magnetic field.}
    \label{f:cdf_toy}
\end{figure}

Further we look at the observational dependence between bulk temperature and magnetic field. We use a fit for surface temperatures similar to work by \cite{Mong18,zelati}:
\begin{equation}
T \propto \alpha B^{0.4},   
\end{equation}
where $T$ stands for surface temperature of the magnetar, $B$ is the initial magnetic field. Using the data collected in the article \cite{Mong18} for 11 observed magnetars in the quiescent state, we obtain an approximation:
\begin{equation}
T = 4.3\times 10^{-7} \left(\frac{B_0}{1\; \mathrm{G}}\right)^{0.4}\; \mathrm{keV}.
\label{e:t}
\end{equation}
This approximation is shown in Figure~\ref{f:kTB}.
Then we use this dependence for temperature and compute luminosity and flux as:
\begin{equation}
L = 4\pi R^2 \sigma T^4 \;\hspace{1cm} \;\; S = \frac{L}{4\pi D^2}, 
\end{equation}
where $D$ is the distance of the magnetar from Earth. We obtain this parameter from the pulsar population synthesis.
In this model we overproduce bright X-ray sources, see Figure~\ref{f:cdf_toy} for toy population synthesis and Figure~\ref{f:cdfBC_nocooling} for more realistic population synthesis when we take fast magnetic field decay into account. The brightest sources produced in model A should be limited by $S\approx 10^{-12}$~erg~cm$^{-2}$~s$^{-1}$. In our model we get plenty of sources with fluxes $10^{-10}$~--~$10^{-11}$~erg~cm$^{-2}$~s$^{-1}$.
It is also interesting to note that if we introduce a magnetic field decay and substitute the instant value of the magnetic field in  eq.~(\ref{e:t}) it is possible to reproduce decay in luminosity. However, in this case, the luminosity distribution starts depending on the timescale of magnetic field decay (and impurity parameter $Q$ respectively) which is not seen in models produced by \cite{Gullon-2015}. Instead, their models for different time scales of magnetic field decay produce flux distributions that are basically identical. 

Therefore a realistic model for luminosity should include both components: (1) spread in initial temperatures as a function of the initial magnetic field and (2) cooling. Our combined model is:
$$
L (t) = 2.5\times 10^3 \pi R^2 \sigma \left(\frac{B_0}{1\; \mathrm{G}}\right)^{1.6} \left(\frac{t}{1\; \mathrm{year}}\right)^{-0.26} \hspace{2cm}
$$
\begin{equation}
\hspace{5cm}\times\exp \left(-\frac{t}{10^5\; \mathrm{years}}\right) \; \mathrm{egs/s}.
\label{e:lum}
\end{equation}
The luminosity evolution for different values of the initial magnetic field for this model is shown in Figure~\ref{f:lumdec}. We see that different tracks enclose all observed magnetars. We substitute this model in our toy population synthesis and show results in Figure~\ref{f:cdf_toy}. As it is expected model with cooling lies well below the actual magnetars which is in agreement with curve for model A in \cite{Gullon-2015}. We do not overproduce bright sources anymore. Further, we use this luminosity model as a basic model in our advanced population synthesis.

The parameters which we selected in eq.~(\ref{e:lum}) are not arbitrary. The value of the exponent is restricted on one hand by a necessity to include the brightest magnetars (e.g. a value of $-0.5$ does not work already) and to have some luminosity decline compatible with simulations, see figure 11 in \cite{vigano2013}. The time scale for luminosity decay cannot be any shorter (we miss magnetars) and cannot be significantly longer (we overproduce bright sources).

\subsubsection{Advanced magnetar population synthesis}
\label{s:realistic_magnetars}
The magnetar population synthesis proceeds in the following steps: (1) we select from the results of pulsar population synthesis NSs with ages $T < 1\times 10^6$~years, with initial magnetic fields above $10^{13}$~G. In this selection we include  objects independently of how probable it is to detect them in a pulsar radio survey; 
(2) we compute their magnetic fields at the current moment taking into account the impurity parameter $Q$; 
(3) we compute their expected temperature as a function of their initial magnetic field and 
(4) we model their X-ray spectra and fluxes taking absorption and distance into account. In the end, we compare the cumulative distribution for X-ray fluxes and final spin periods. 

\subsubsection{Magnetic field decay}

We use the phenomenological model described in \cite{2015AN....336..831I,2018MNRAS.473.3204I,igoshevpopov2020,igoshevpopov20201}. Namely, we integrate numerically an approximate differential equation by \cite{aguilera2008}:
\begin{equation}
\frac{dB_p}{dt} = - \frac{B_p}{\tau_\mathrm{Ohm}(T_\mathrm{crust})} - \frac{1}{B_0} \frac{B_p^2}{\tau_\mathrm{Hall}},    
\end{equation}
where $B_p$ is the poloidal, dipolar magnetic field at the NS pole, $B_0$ is the initial poloidal dipolar field strength, $\tau_\mathrm{Hall}$ is the initial Hall timescale and $\tau_\mathrm{Ohm}$ is the Ohmic timescale for the temperature of the deep NS crust $T_\mathrm{crust}$. 
One of the main factors which govern the magnetic field evolution for magnetars is their crust impurity. We relate the timescale for magnetic field decay with a value of the impurity parameter as:
\begin{equation}
\tau_{Q} = \frac{2\; \mathrm{Myr}}{Q},   
\label{e:q_decay}
\end{equation}
which roughly corresponds to the case when most of the current flow at densities $\approx 5\times 10^{13}$~g~cm$^{-3}$. Based on magnetic field decay curve figure 2 by \cite{gullon2014} we see that their relationship between time scale and $Q$ is:
\begin{equation}
\tau_Q \approx \frac{20\; \mathrm{Myr}}{Q},    
\end{equation}
because they assume that most of the current flows at densities $\approx 5\times 10^{14}$~g~cm$^{-3}$. Therefore their $Q = 100$ roughly corresponds to our $Q = 10$ and their $Q = 25$ would correspond to our $Q=2.5$.
This difference is simply related to a different choice of parameters and can be easily re-scaled in our model if new observations prove that our choice is nonphysical.  

\begin{figure}
	\includegraphics[width=\columnwidth]{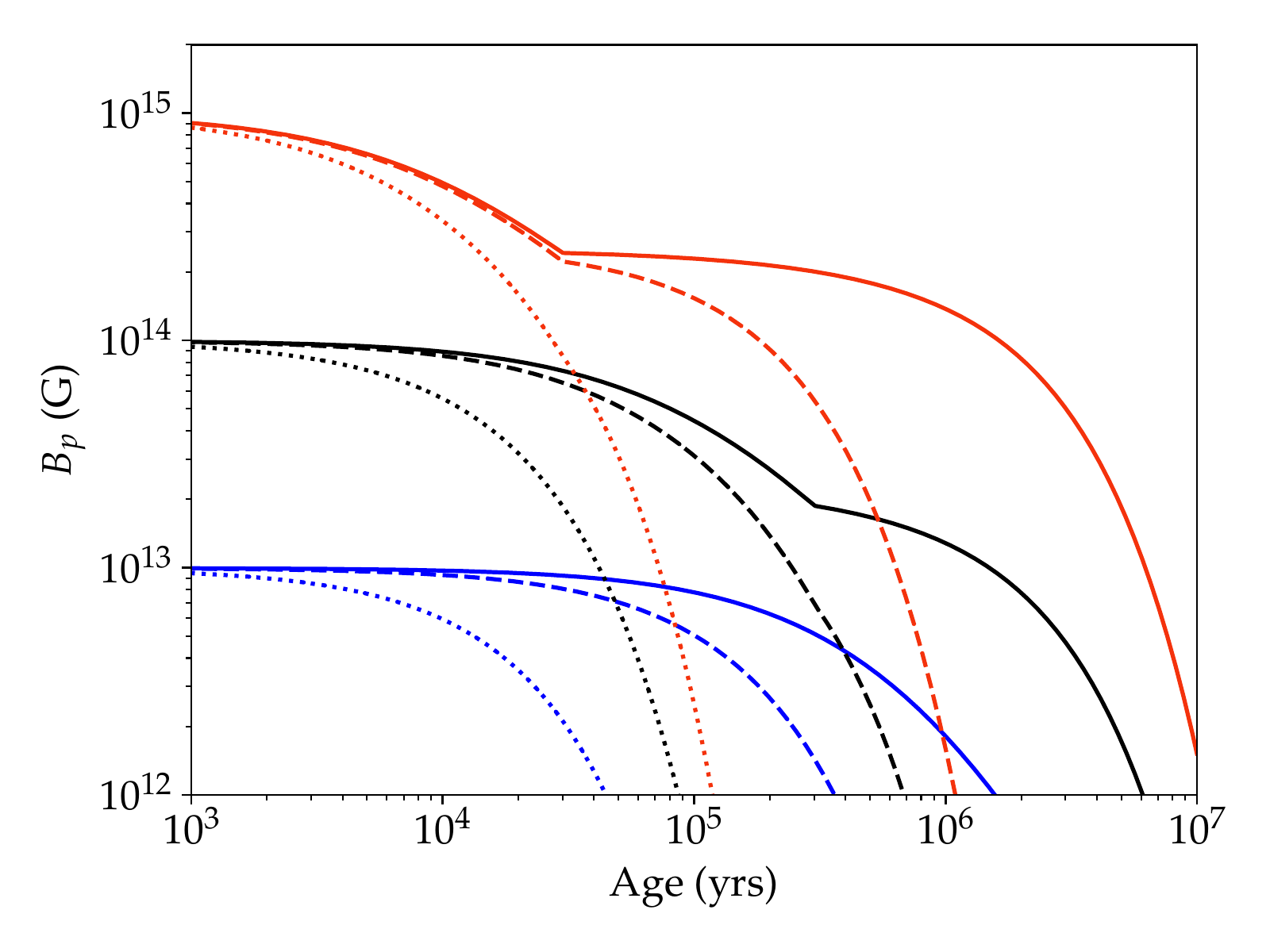}
\caption{Strength of the poloidal dipolar magnetic field as a function of time and impurity parameter $Q$. Solid lines correspond to $Q = 1$, dashed lines to $Q = 10$ and dotted lines to $Q = 100$.}
    \label{f:mf_evolv}
\end{figure}

\begin{figure}
\center{\includegraphics[width=\columnwidth]{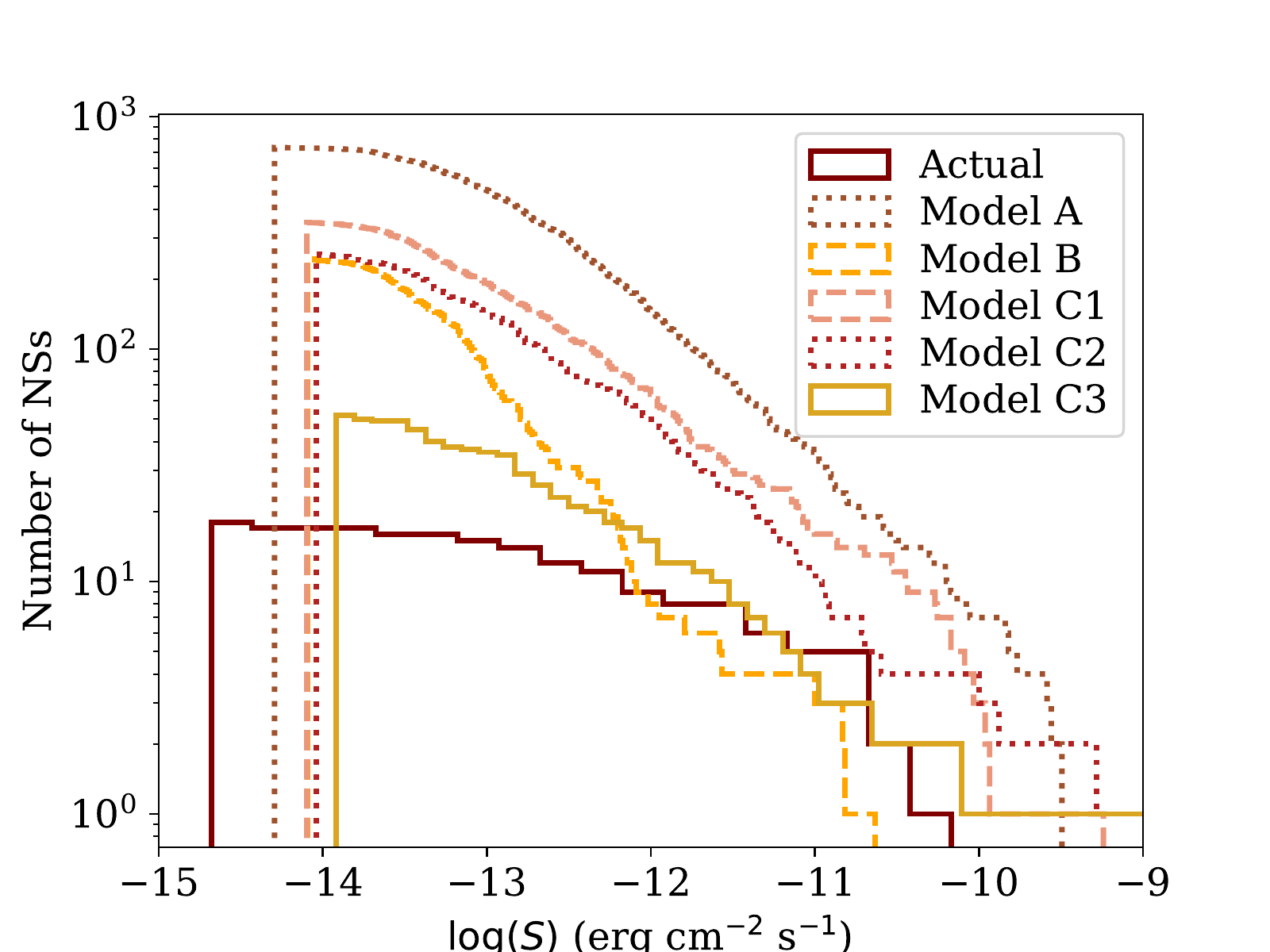}}
\caption{Cumulative distribution function for X-ray fluxes for magnetars. Maroon line is real observations from \protect\cite{Mong18}, see Table \ref{tab:res} for details of each model. X-ray luminosities decay only due to decay of surface dipolar poloidal magnetic field, see Section~\ref{s:toy_xray} for details.}
\label{f:cdfBC_nocooling}
\end{figure}

We show our magnetic field decay curves in Figure~\ref{f:mf_evolv}. As it was discussed in Section~\ref{s:toy_xray}, the magnetic field decay law does not affect the X-ray flux, so it affects only the distribution of observed periods.

\subsubsection{Modelling spectra and observational X-ray fluxes}
\label{s:xray_spectra}

Observations of magnetars in the quiescent state showed that their spectra consist both of thermal blackbody emission and non-thermal power-law tail at higher energies. We use the model by \cite{Lyutikov} to take into account the scattering of thermal photons on the electrons of the magnetosphere. It is parameterized by two quantities: the optical depth $\tau_\mathrm{res}$ (which is related to the plasma density) and the thermal velocity of electrons in units of the speed of light $\beta_\mathrm{T}$. The corrected formulas of the reflected and transmitted flux are given in Appendix \ref{A}. 

Further, we correct these spectra for interstellar absorption. For the case of interstellar absorption an effective model of resonant Compton scattering could improve the number of detectable NSs, because the Compton scattering shifts low-energy photons to higher energies, so they can avoid absorption.

To account for absorption we use the Leiden/Argentine/Bonn (LAB) Survey of Galactic HI sky \cite{LAB_map}. The final spectrum is then \citep{Potekhin16}:
\begin{equation}
    F_\mathrm{fin} = F(\omega)\cdot \exp\left[-\left(\frac{N_H}{10^{21}\mathrm{cm^{-2}}}\right) \cdot \left( \frac{h \omega}{2 \pi\;  0.16\; \mathrm{keV}} \right)^{-8/3}\right],
\end{equation}
for each step of energy in the range from $0.1$ to $10$~keV.
Here $F(\omega)$ is the flux before absorption, $N_H$ is the hydrogen density column in the direction of the magnetar, $h$ is the Planck constant and $\omega$ is the radiation frequency.\\
By combining the spectrum with a map of Galactic hydrogen and normalising it, we compute how much of the radiation was absorbed (denote this value as $\eta$, $ \mathrm{\eta} \geq$  1). 
\begin{equation}
\eta = \frac{\int F_\mathrm{fin} (\omega)d\omega}{\int F_\mathrm{bb}(\omega)d\omega}.    
\end{equation}
Where we integrate from 0.1~keV to 10~keV.
As a result we compute the received X-ray flux as:
\begin{equation}
S = \frac{L_\mathrm{bol}}{4 \pi D^2} \cdot \mathrm{\eta}.
\end{equation}
We assume that the thermal radiation from magnetars is not beamed. 





\section{Results}
\label{s:res}

\begin{figure}
\center{\includegraphics[width=\columnwidth]{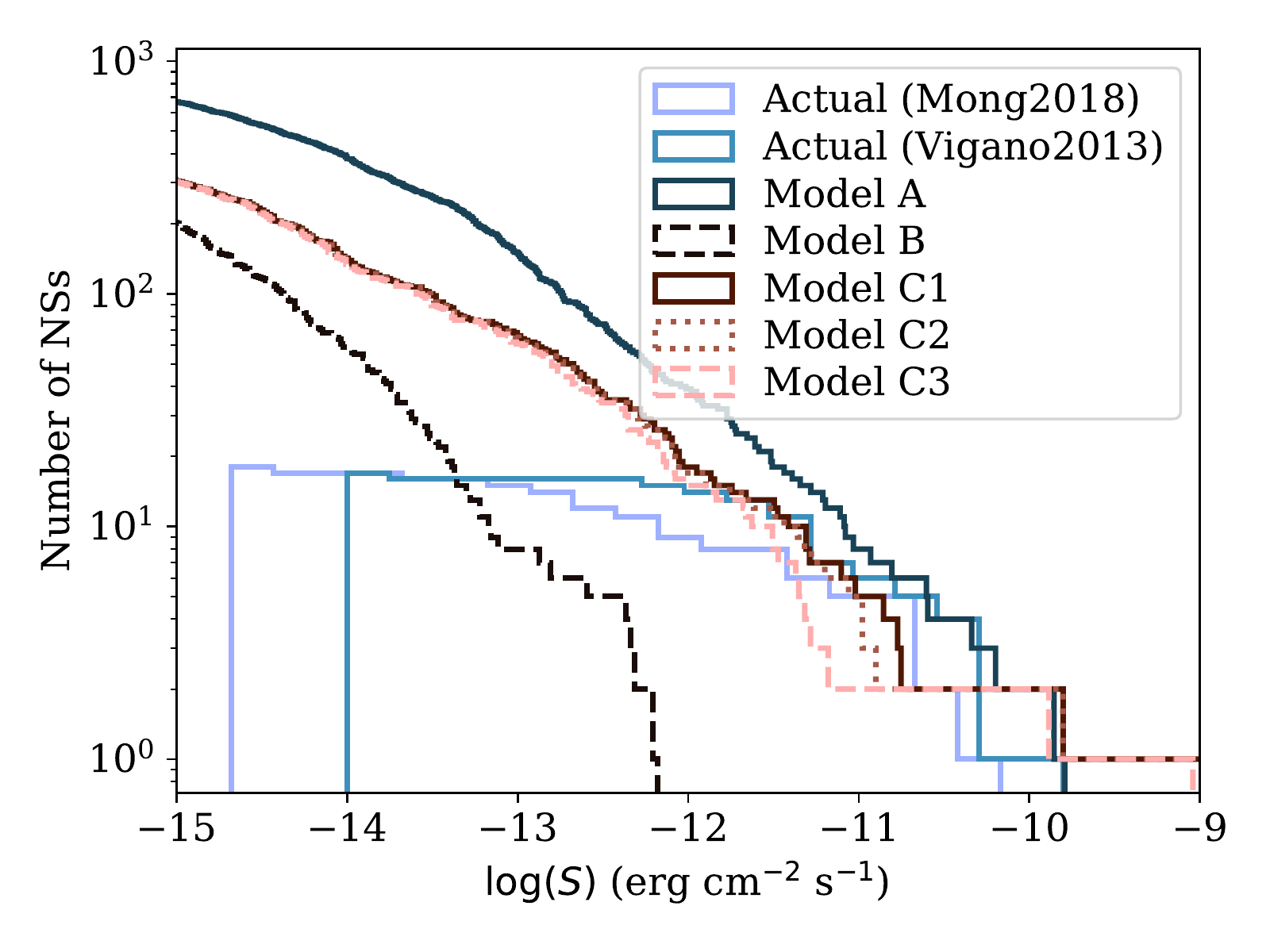}}
\caption{Cumulative distribution for X-ray fluxes of magnetars. Light blue line shows observations by \protect\cite{Mong18}; blue line shows observations by \protect\cite{vigano2013}. See Table \protect\ref{tab:res} for details of each model.}
\label{f:cdfABC}
\end{figure}

\begin{figure}
\center{\includegraphics[width=\columnwidth]{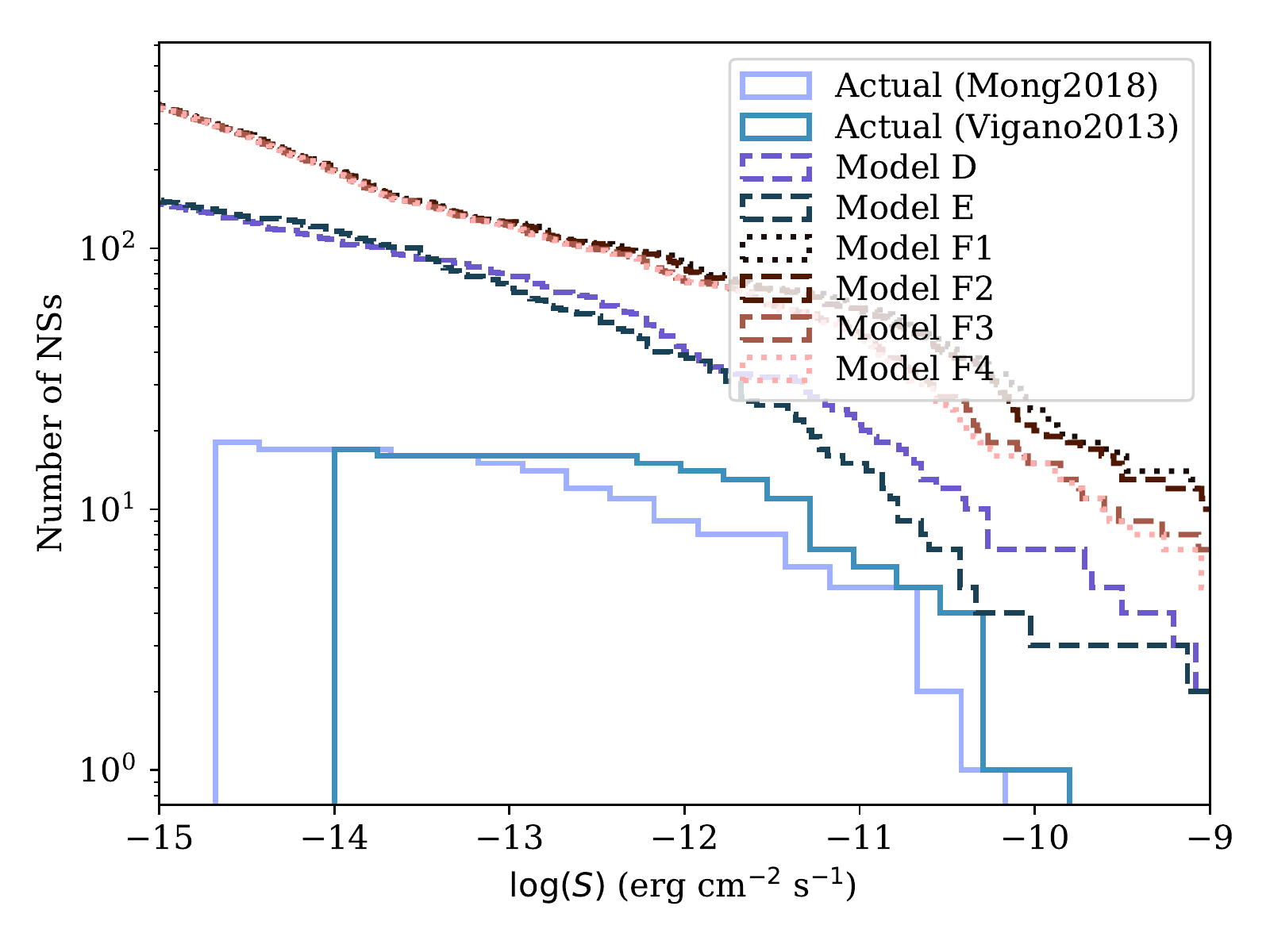}}
\caption{Cumulative distribution function for magnetars. Real observations are from \protect\cite{Mong18}, Model E is our approximation that initial magnetic filed should be two gaussians with weakly and stongly magnetic stars parameters, 
inherited from massive stars, see \ref{tab:res} for details of each model.}
\label{f:cdfEF}
\end{figure}

We summarise the results of our runs in Table~\ref{tab:res}. We also plot the cumulative distribution of X-ray fluxes in Figure~\ref{f:cdfABC} and \ref{f:cdfEF}. We notice that our run with initial parameters similar to \cite{2006ApJ...643..332F} reproduces well the population of isolated radio pulsars as it is expected, see results for model B and Figure~\ref{f:pdotpAB}. This model does not describe the population of magnetars, because it does not produce enough stars with magnetic fields above $4\times 10^{13}$~G, so we see a lack of magnetars with observed X-ray fluxes in the range $10^{-10}$~--~$10^{-12}$~erg~cm$^{-1}$~s$^{-1}$, see Figure~\ref{f:cdfABC}. This result is in agreement with \cite{Gullon-2015}. Model A with a slightly larger initial magnetic field overproduces the strongly magnetised radio pulsars, see Figure~\ref{f:pdotpAB}. This is why its C-statistics value is larger in comparison to model B.

\begin{table*}
    \centering
    \begin{tabular}{ccccccccll}
    \hline
    Model & $w$  & $\mu_{B,1}$ & $\sigma_{B,1}$ & $\mu_{B,2}$ & $\sigma_{B,2}$ & $Q$ & C-stat &  Comment \\
          & \% & $\log_{10} $ G &             & $\log_{10} $ G &             &  & per d.o.f  & \\
    \hline
    A     & 100 & 13.34 & 0.76  & &            & 1 & 16.6 &  similar to Model B (Table 2) in \cite{Gullon-2015} \\
    B     & 100 & 12.65 & 0.55  & &            & 1 & 7.89 &  similar to \cite{2006ApJ...643..332F} \\
    C1    & 70  & 12.59 & 0.59  & 13.33 & 0.83 & 1   & 6.61 & similar to Model F (Table 3,4) in \cite{Gullon-2015} \\
    C2    & 70  & 12.59 & 0.59  & 13.33 & 0.83 & 10  & 6.61 & similar to Model F (Table 3,4) in \cite{Gullon-2015} \\
    C3    & 70  & 12.59 & 0.59  & 13.33 & 0.83 & 100 & 6.61 & similar to Model F (Table 3,4) in \cite{Gullon-2015} \\
    D     & 90 & 12.2 & 0.6 & 14.7  & 0.6 & 1 & 15.2 & simple flux conservation eq.(\ref{e:flux_conservation}) \\
    E     & 90  & 11.7  & 0.7   & 14.45 & 0.7  & 1   & 38.64  & simple model for core conductivity eq.(\ref{e:core_cond}) \\
    F1    & 90  & 12.65 & 0.7   & 15.35 & 0.7  & 1   & 7.59 & fossil field hypothesis \\
    F2    & 90  & 12.65 & 0.7   & 15.35 & 0.7  & 10   & 7.59 & fossil field hypothesis \\
    F3    & 90  & 12.65 & 0.7   & 15.35 & 0.7  & 100  & 7.59 & fossil field hypothesis \\
    F4    & 90  & 12.65 & 0.7   & 15.35 & 0.7  & 200  & 7.59 & fossil field hypothesis \\
    \hline
    \end{tabular}
    \caption{Models calculated in our research and their results. C-stat value is divided by the total number of bins (20$^{2}$). $w$ shows a fraction of objects drawn from the first component of bimodal log-normal distribution.  
    }
    \label{tab:res}
\end{table*}

\begin{figure}
	\includegraphics[width=\columnwidth]{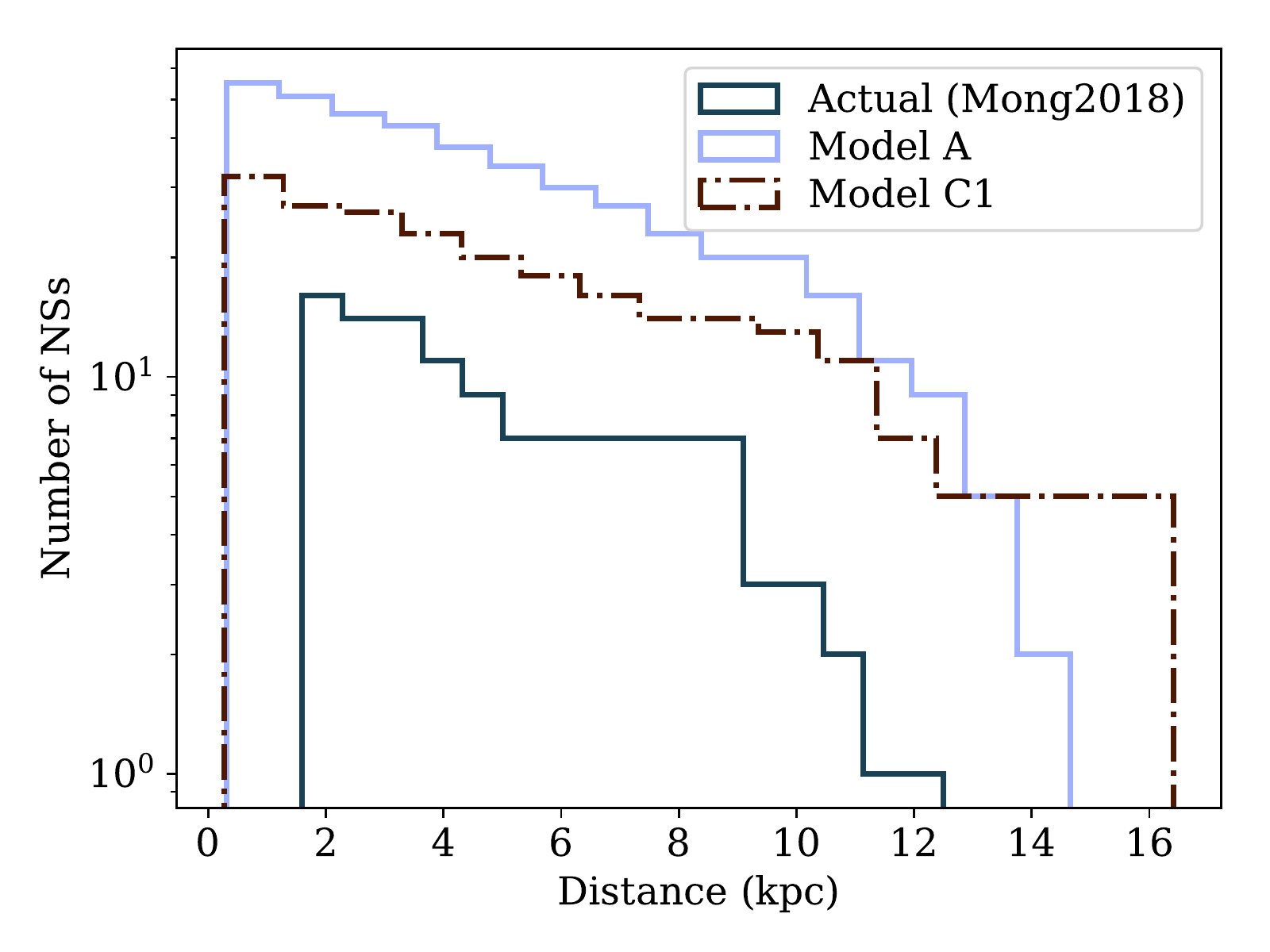}
\caption{Cumulative distribution function for magnetars with fluxes $> 5 \cdot 10^{13}$  $\mathrm{erg cm^{-2} s^{-1}}$. See Table \ref{tab:res} for details about each model. Two magnetars are excluded from the actual sample (SGR 0526--66 in the LMC and CXOU J010043.1--721134 in the SMC), because we are considering only the galactic population.}
    \label{f:cdf_Dist_ABC}
\end{figure}

\begin{figure}
\center{\includegraphics[width=\columnwidth]{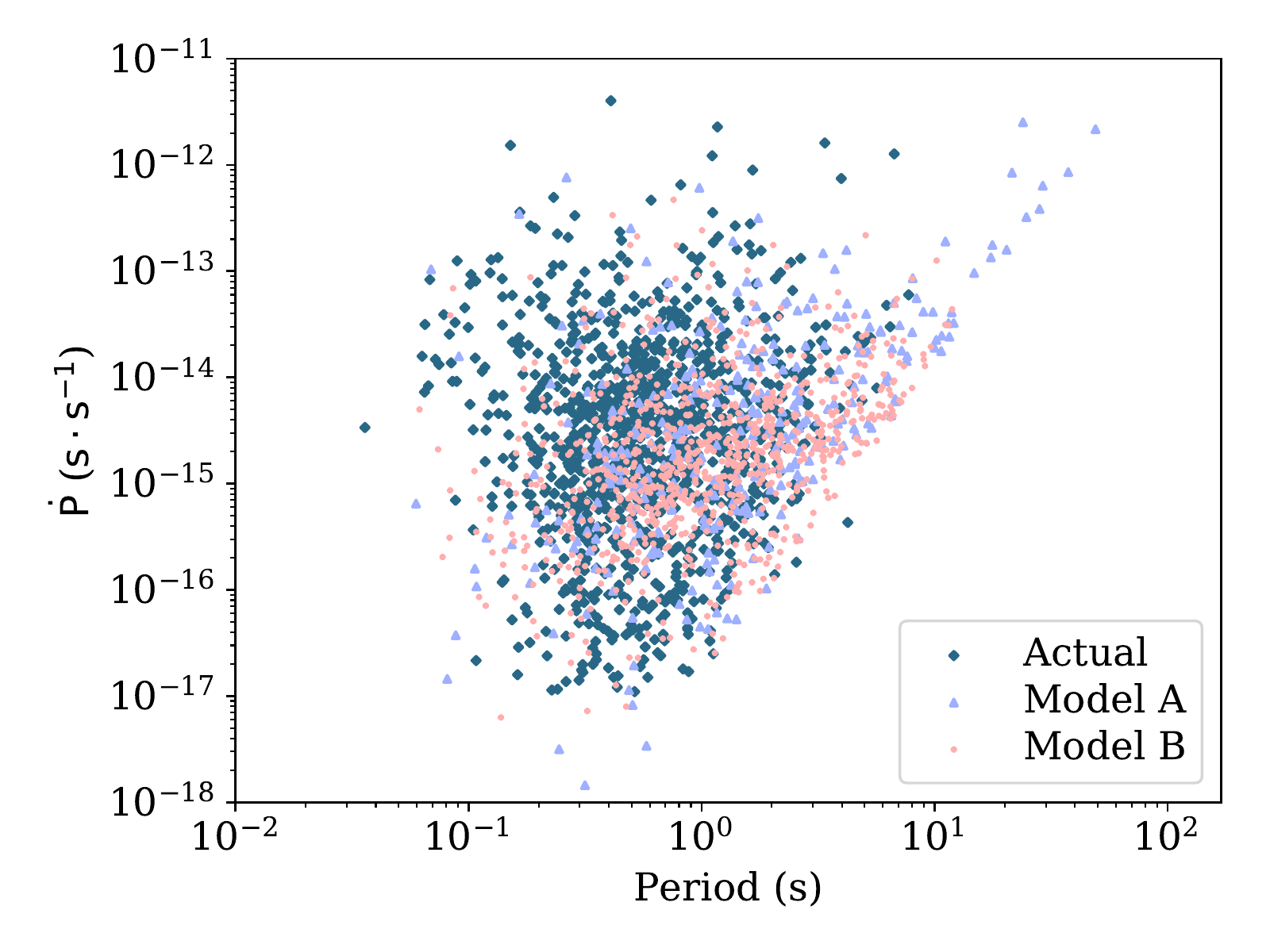}}
\caption{$P$ - $\dot{P}$ diagram for real pulsars and for Model A, B.}
\label{f:pdotpAB}
\end{figure}

\begin{figure}
\center{\includegraphics[width=\columnwidth]{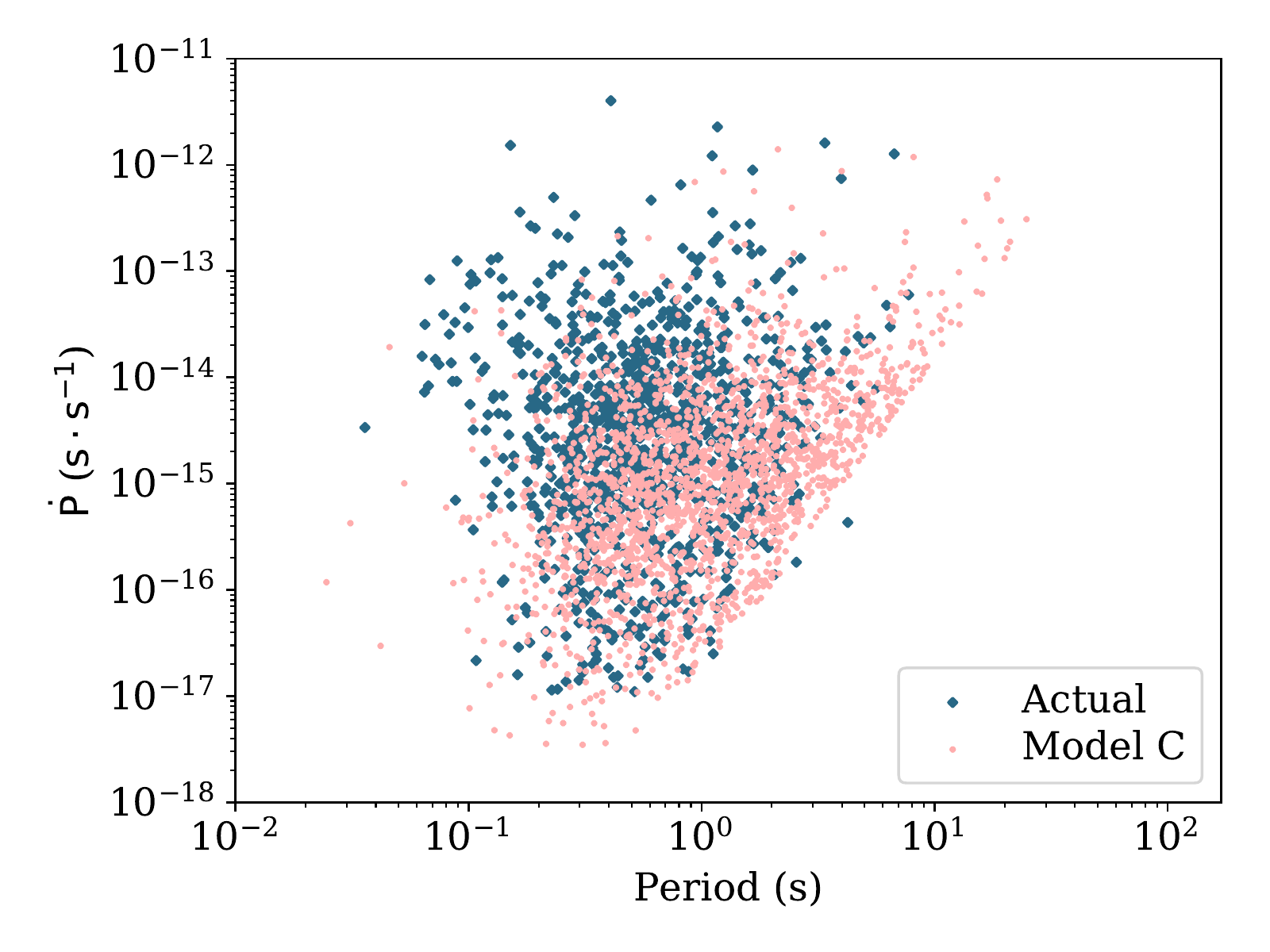}}
\caption{$P$ - $\dot{P}$ diagram for real pulsars and for Model C.}
\label{f:pdotpC}
\end{figure}

Models C1-C3 with initial parameters from \cite{Gullon-2015} could reproduce the population of isolated radio pulsars quite well, see Figure~\ref{f:pdotpC}. The value of C-stat per degree of freedom is even slightly smaller than in the case of the initial conditions similar to \cite{2006ApJ...643..332F}. As for the magnetars, all C models reproduce the observed X-ray fluxes well, see Figure~\ref{f:cdfABC}, taking into account the fact that our current catalogue of magnetars is probably incomplete below fluxes $5\times 10^{-13}$~erg~cm$^{-2}$~s$^{-1}$. A similar threshold value was also noted by \cite{Gullon-2015}. It is also interesting to note that if we select only magnetars with fluxes above this threshold, we closely reproduce the magnetar distance distribution, see Figure~\ref{f:cdf_Dist_ABC} (with exception of model A). 

As for observed period distribution, we would prefer impurity values between $10$ and $100$ because C2 overproduces magnetars with observed periods longer than 12~s and model C3 under-produces a total number of magnetars, see Figure~\ref{f:pBC} 
It is possible to find the exact value of the impurity parameter $Q\in [10,100]$ which describes the period distribution of magnetars.

\begin{figure}
\center{\includegraphics[width=\columnwidth]{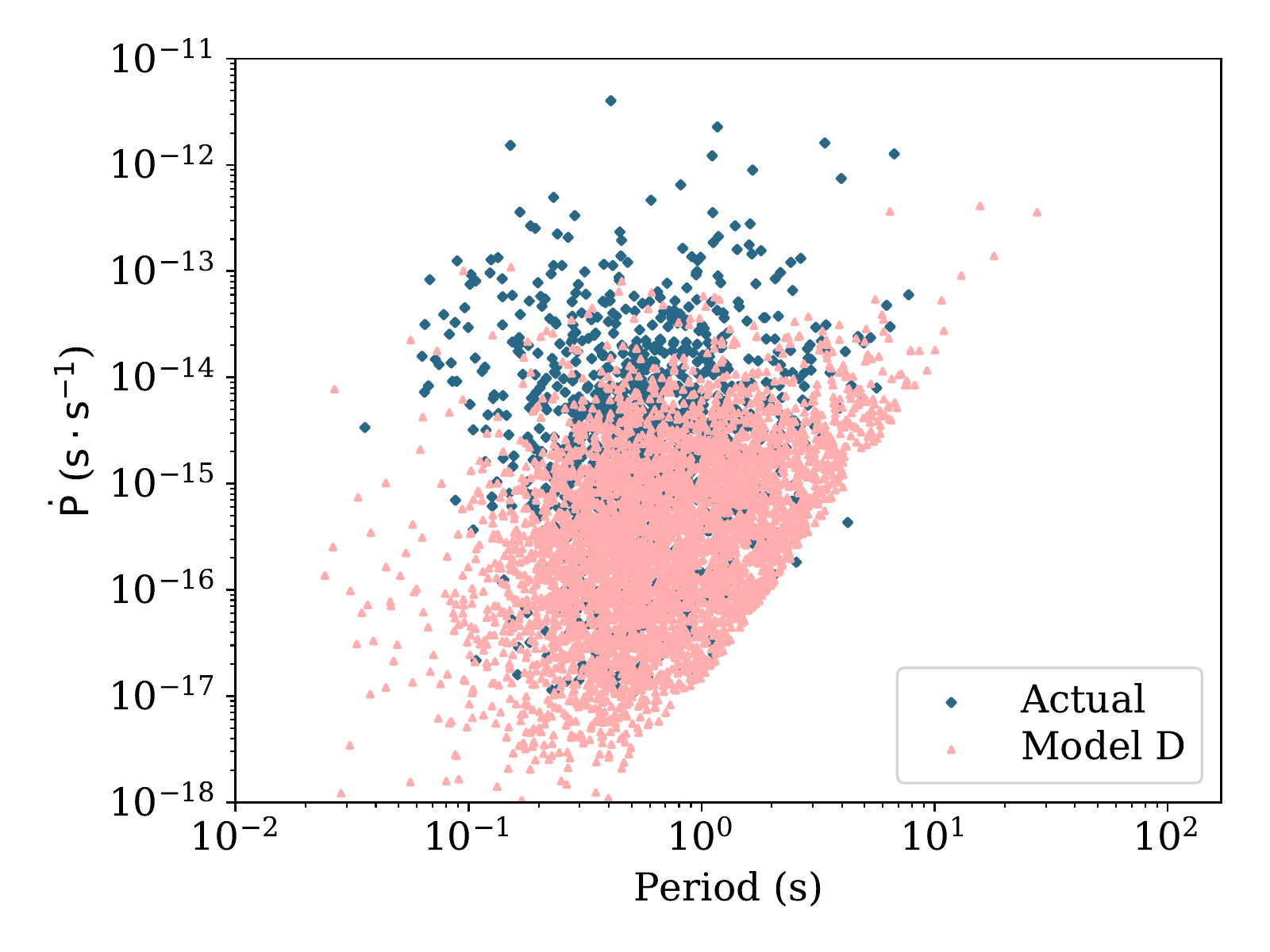}}
\caption{$P$ - $\dot{P}$ diagram for real pulsars and for Model D.}
\label{f:pdotpD}
\end{figure}

To test the fossil field hypothesis, we first introduce model D where we compute the initial distribution of NS magnetic fields using simple flux conservation for O stars, see eq. (\ref{e:flux_conservation}). As it is seen from $P$~--~$\dot P$ plot, Figure~\ref{f:pdotpD}, the mean magnetic field $\mu_B = 12.2$ is clearly small for normal radio pulsars. We produce too many weakly magnetised pulsars in this case. On the other hand, we produce too many bright magnetars (approximately 10) with $S_X$ in range $10^{-8}-10^{-10}$~erg~cm$^{-2}$~s$^{-1}$, see Figure~\ref{f:cdfEF}. Such magnetars are not observed.

\begin{figure}
\center{\includegraphics[width=\columnwidth]{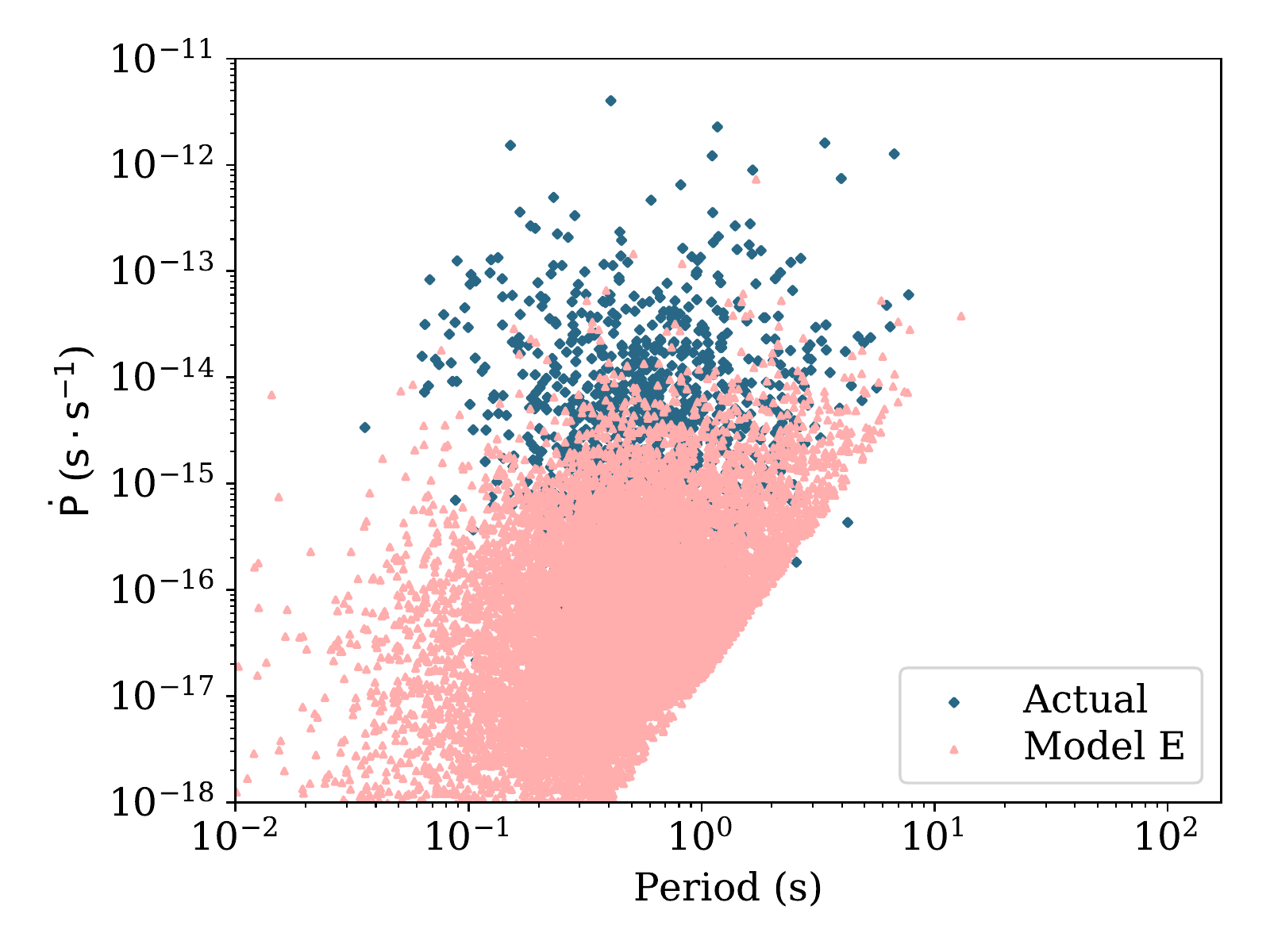}}
\caption{$P$ - $\dot{P}$ diagram for real pulsars and for Model E.}
\label{f:pdotpE}
\end{figure}

Further, we introduce model E based on our simple estimates of magnetic field inherited to NS at the moment of a supernova explosion, described in Section~\ref{s:core_conduct}. We notice that initial magnetic fields of radio pulsars produced in this model $\mu_B = 11.7$ are smaller than typical for observations ($\mu_B = 12.6$ in \citealt{2006ApJ...643..332F}) by nearly an order of magnitude. The value of C-statistics and visual inspection of Figure~\ref{f:pdotpE} immediately show that the main pulsar cloud is significantly shifted toward small period derivative values. Therefore, we did not try to optimise this model by using different values of impurity parameter $Q$. 

\begin{figure}
\center{\includegraphics[width=\columnwidth]{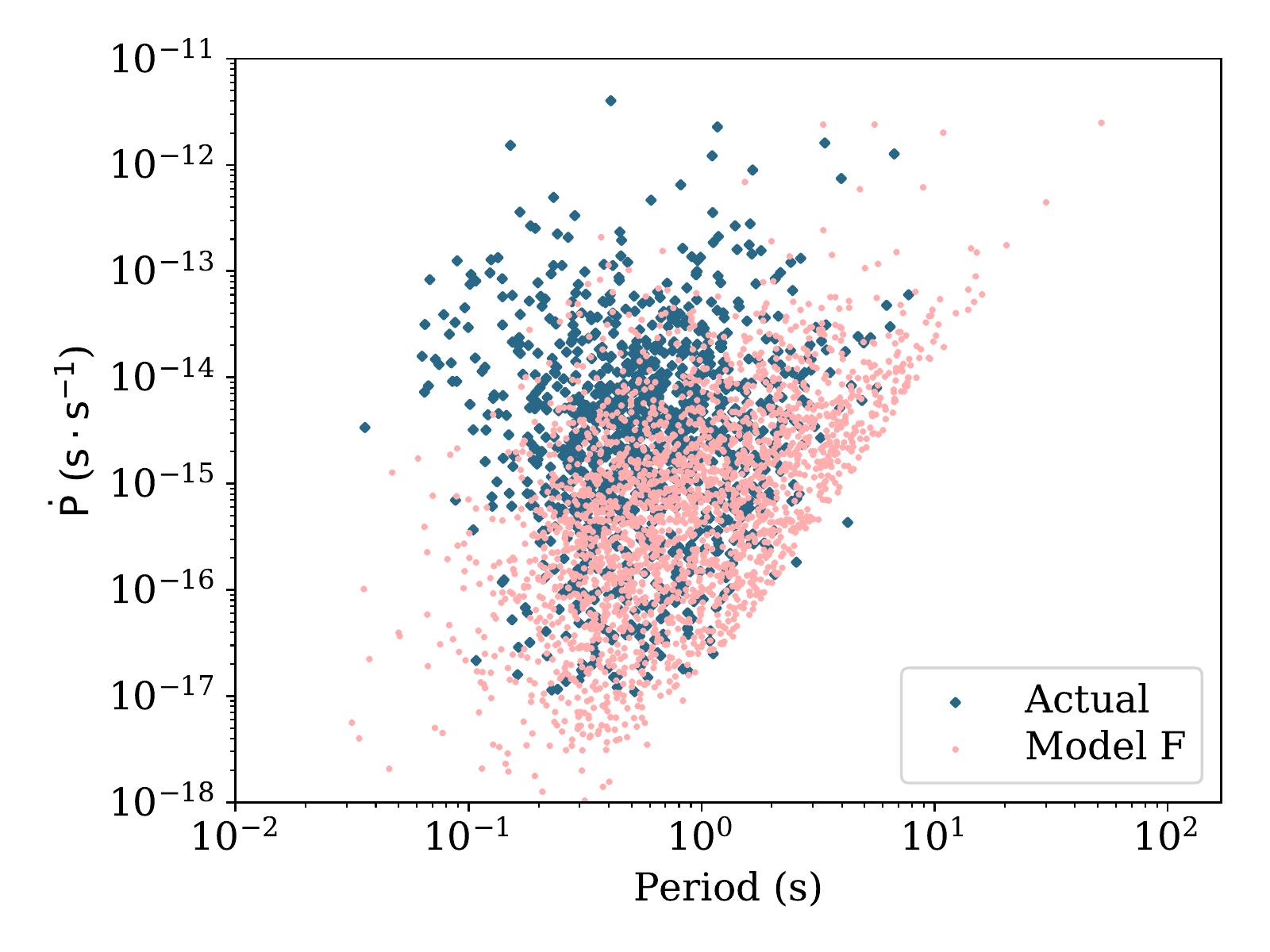}}
\caption{$P$ - $\dot{P}$ diagram for real pulsars and for Model F.}
\label{f:pdotpF}
\end{figure}

Here it is important to notice that $\mu_B = 2.83$ of strongly magnetised B stars  differs from $\mu_B = 0.15$ of weakly magnetised stars by $\approx 2.7$ DEX. It means that independently of our model for magnetic field conservation, the initial magnetic fields of magnetars will be 2.7 DEX stronger than magnetic fields of radio pulsars. \cite{Gullon-2015} obtained a difference of 0.68-1.1 DEX between $\mu_B$ for pulsars and magnetars. The only way to reconcile such a difference is to assume a strong magnetic field decay for magnetars, possibly even stronger than \cite{Gullon-2015} assumed. Moreover, when \cite{Gullon-2015} tried to fit a model with bimodal magnetic field distribution, they attributed 30-40\% of all NSs to a strongly magnetised group, which is significantly more than 7-12\% of massive stars being strongly magnetic. 

\begin{figure}
\center{\includegraphics[width=\columnwidth]{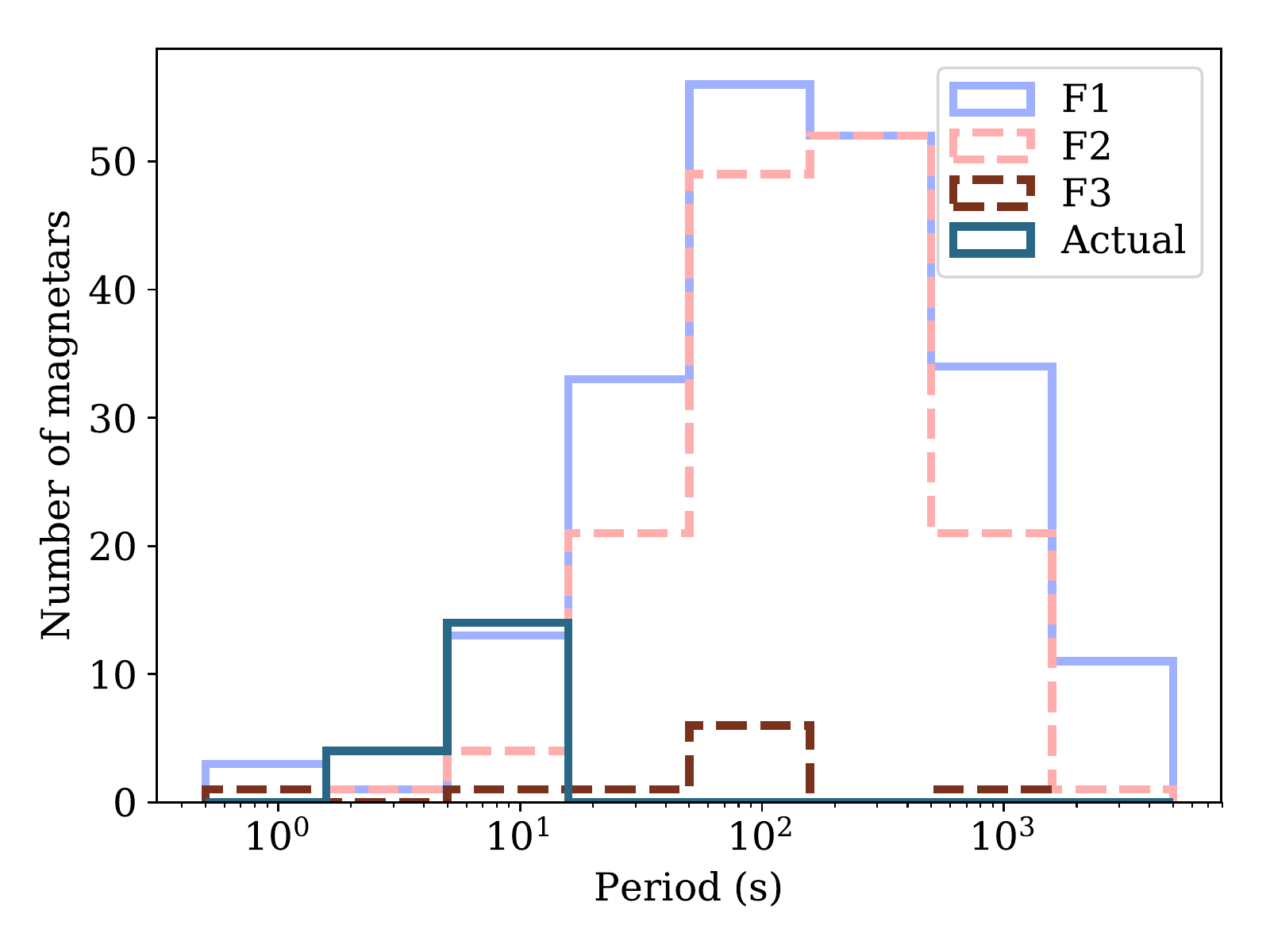}}
\caption{Distribution of periods for magnetars: Model F1, F2, F3.}
\label{f:pEF}
\end{figure}

\begin{figure}
\center{\includegraphics[width=\columnwidth]{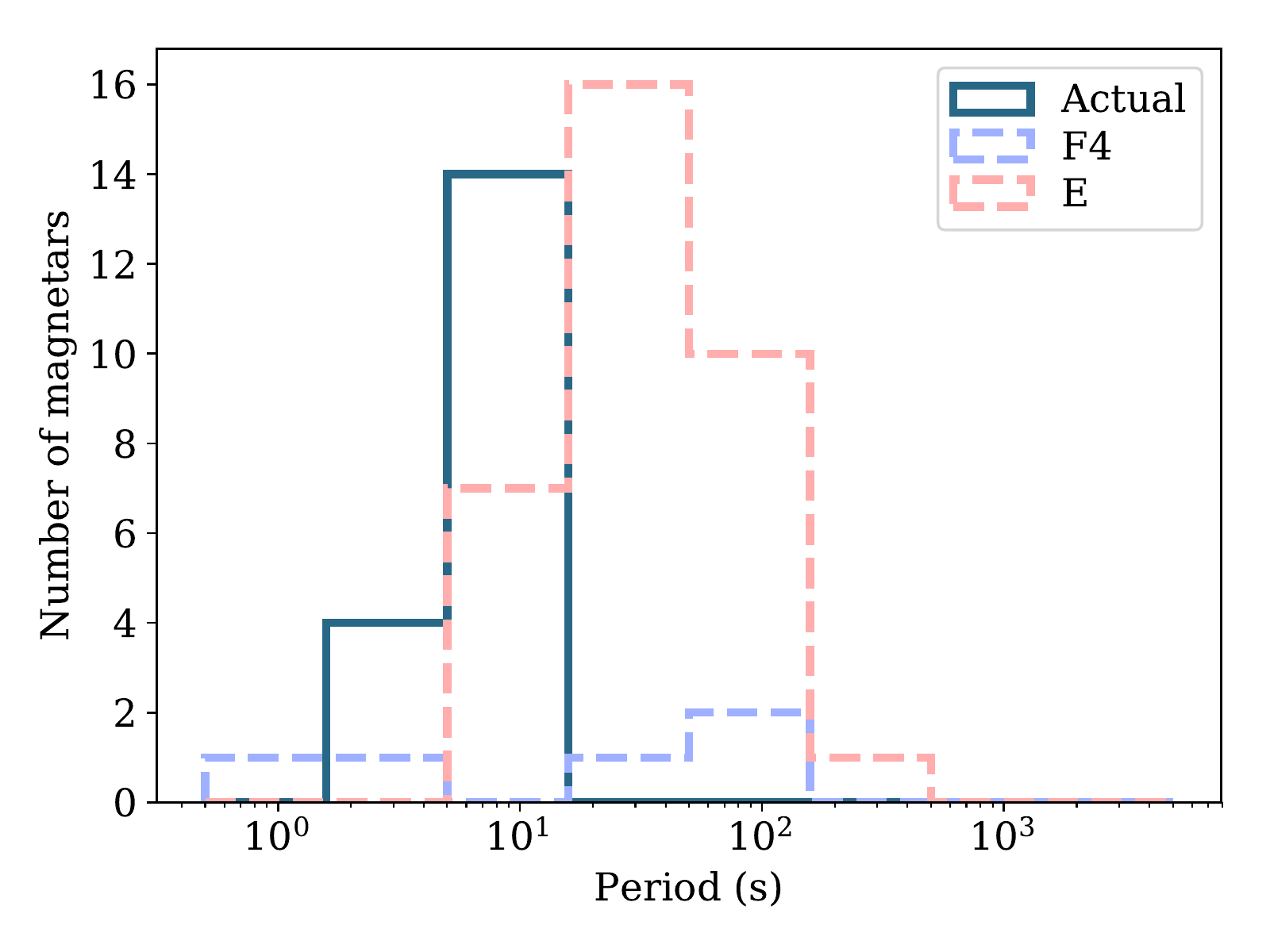}}
\caption{Distribution of periods for magnetars: Model F4, E.}
\label{f:pEF2}
\end{figure}

\begin{figure}
\center{\includegraphics[width=\columnwidth]{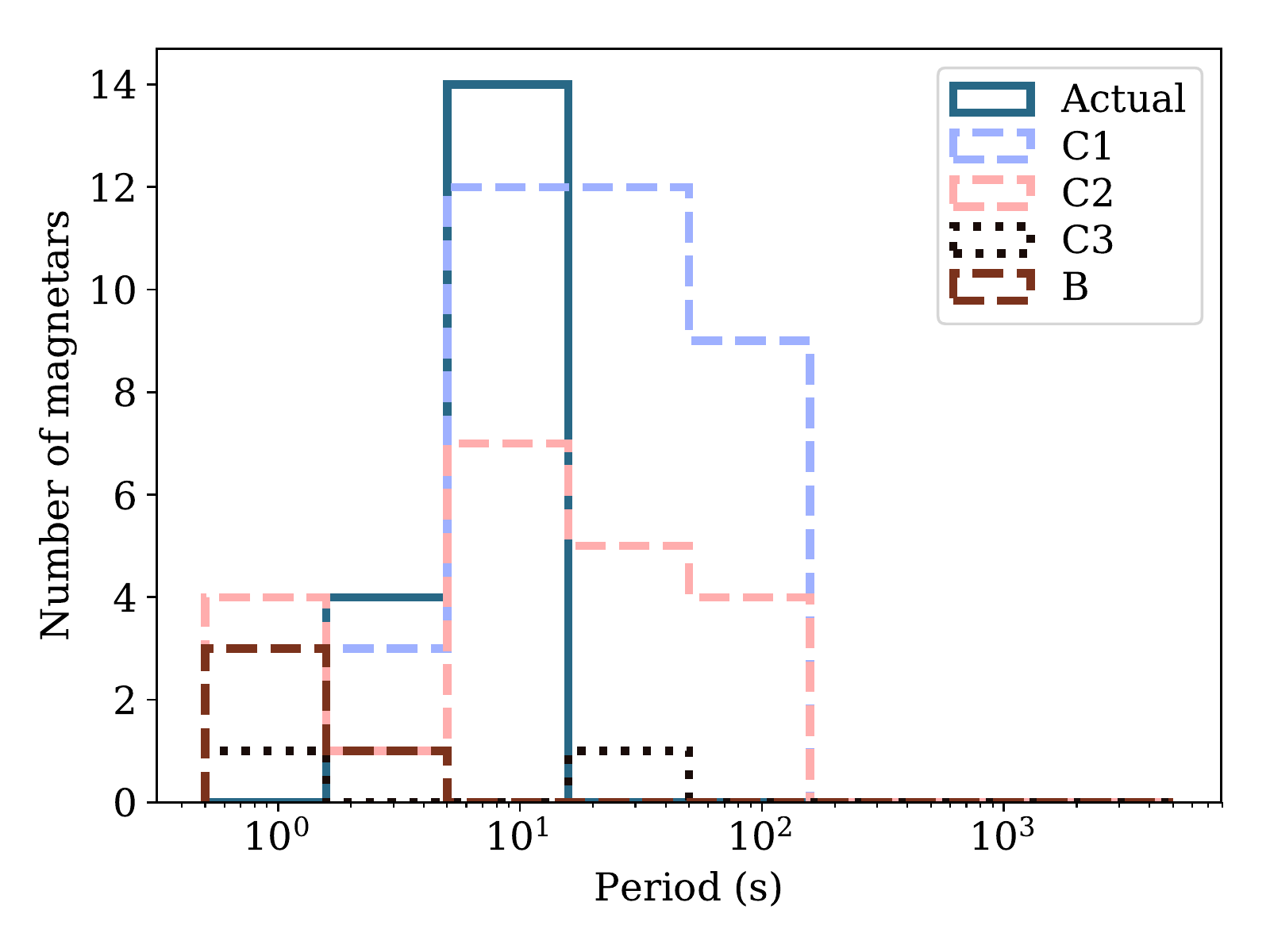}}
\caption{Distribution of periods for magnetars: Models B and C1, C2, C3. Real data is also presented.}
\label{f:pBC}
\end{figure}

To check if fast magnetic field decay could help with this discrepancy we computed models F1-F4 where we set the $\mu_B=12.65$ at a value from the \cite{2006ApJ...643..332F}, so we can describe the population of isolated radio pulsars well and $\mu_B = 14.45$ to satisfy 2.7 DEX difference of magnetic fields between strongly and weakly magnetised B stars.
As it can be seen from the Table~\ref{tab:res}, these models reproduce the isolated radio pulsars reasonably well. Inspection of the period -- period derivative plot Figure~\ref{f:pdotpF} also shows no significant problems. 
These models significantly overproduce number of magnetars with $S_X$ in range $10^{-10}$~--$10^{-8}$~erg~cm$^{-2}$~s$^{-1}$, see Figure~\ref{f:cdfEF}. The impurity parameter cannot help with this problem because we assume that the X-ray luminosity depends only on the initial magnetic field (similar behaviour is seen in detailed simulations by \citealt{Gullon-2015}). Models F1-F3 produce magnetars with periods significantly longer than 12~s, see Figure~\ref{f:pEF}. Models F4 and E give slightly better distribution for periods of magnetars but still produce multiple sources with periods longer than 12~s, see Figure~\ref{f:pEF2}.
Therefore it seems impossible to reconcile the difference in magnetic fields of weakly and strongly magnetised massive stars (2.7 DEX) with a necessity to simultaneously describe normal radio pulsars and magnetars.


\section{Discussion}
\label{s:discussion}
In this section we discuss different properties of magnetised massive stars such as fraction of these stars in binary systems and their rotational velocities. We also discuss different aspects of magnetar evolution and how magnetic fields of NSs could be coupled with magnetic fields of their progenitors. 

\subsection{Fraction of magnetic stars in binary systems}\label{BinFrac}

In our catalogue, we see that among nine  weakly-magnetic stars four are in binaries (mostly spectroscopic). For O stars: $\approx 27\%$ are in binaries; for B stars: $\approx 24\%$. Typical values for non-magnetic OB-stars are $\approx 29\%$ or higher (e.g. \citealt{Aldoretta15} HST all-sky survey of OB-stars or \citealt{Sana17} overview). For strongly magnetised stars, the observational fraction is not well defined and varies greatly depending on the sample and the observational procedure as the fraction is obtained depending on the sample/survey of OB stars. For A star we obtain value $\approx 23\%$ versus $>50\%$ by \cite{Duchene13} for observed non-magnetic and $\approx 23\%$ for observed magnetic AB-stars \cite{Rastegaev14}. At the same time the large catalogue by \cite{Mathys17} of magnetic Ap-stars shows binary percentage around 50\%.

BinaMIcS project \citep{BMIcS} revealed that magnetism is much less present in binaries than it is in massive single stars. Compared to 7-10 percent obtained for 500 single stars in the MiMeS project, no magnetic fields were detected in 700 binaries, where at least one star was of spectral class O, B or A. The detection threshold was the same as in MiMeS. So, it seems that magnetism is less frequent in binary systems. This should be related to the theory of the formation of magnetic fields in multiple systems, but unfortunately, this is also still an open question (it might be a result of the merger \citealt{Schneider2019} or processes during the pre-main sequence \citealt{Villebrun19}).
     
Thus, it is difficult to conclude whether we have a high percentage of binaries or not. First, because there is no clear limit or number in the observational data. Secondly, our sample is subject to strong selection effect due to the complexity of measurements of magnetic fields of massive stars. This is one of the open questions in the evolution of massive magnetic and non-magnetic stars \citep{open_problems2020}, so we will not draw any additional conclusions.

\subsection{Rotation velocity}
\begin{figure}
\center{\includegraphics[width=\columnwidth]{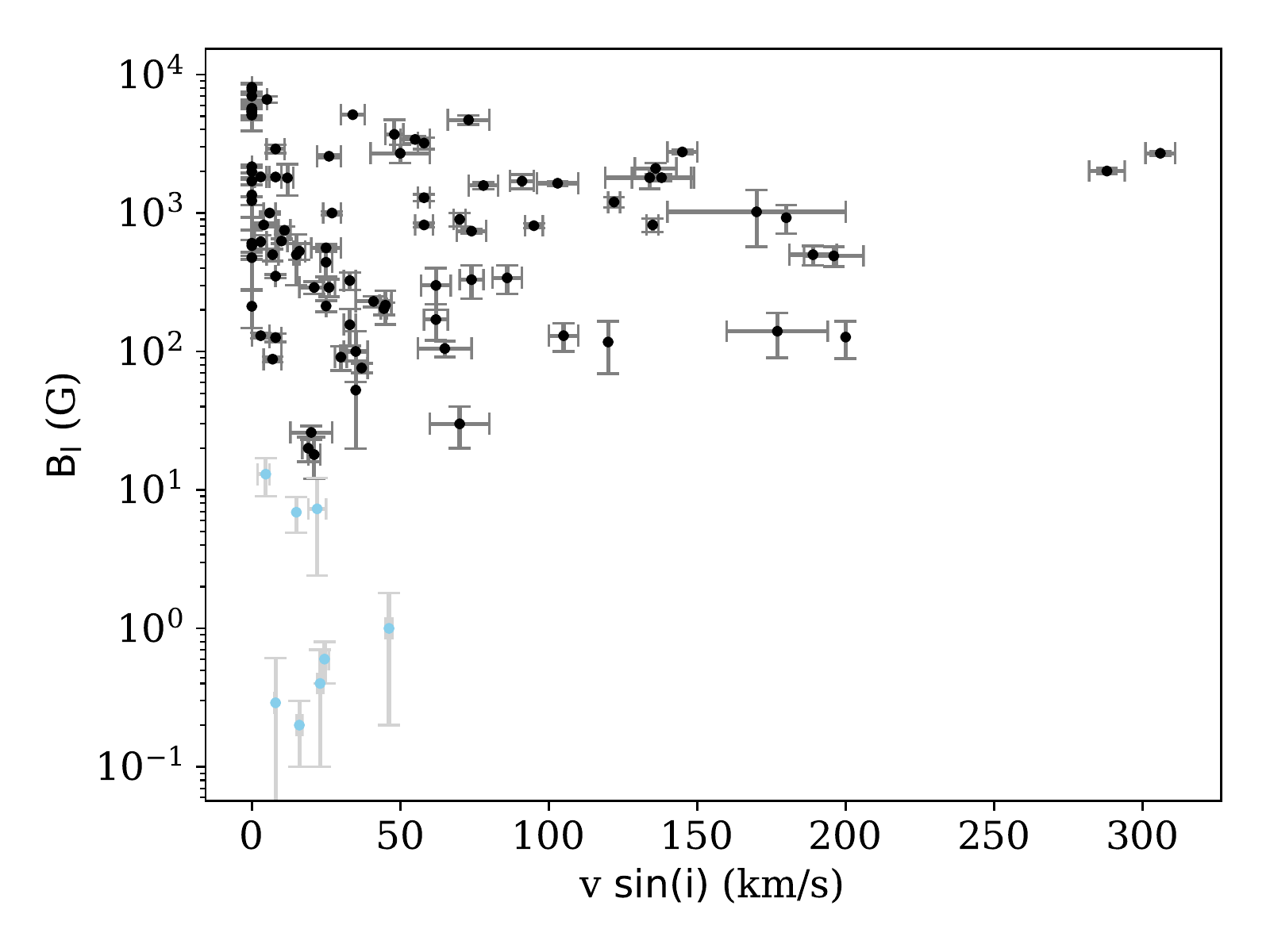}}
\caption{Rotation velocity relative to the magnetic field for B-stars (black) and weakly-magnetic stars (blue).}
\label{f:vsini}
\end{figure}

The stellar evolution models used \citep{2000MNRAS.315..543H} do not take into account stellar rotation. Nevertheless, it is interesting to check if there is a dependence of the rotation velocities on the magnetic field in B-type and weakly-magnetic stars. Large studies of the connection between stellar evolution and rotation have already been carried out by \cite{Brott11} and \cite{Ekstrom12a}. Similar studies were performed taking into account the dependence of the magnetic field of the star and its rotation velocity by \cite{Shultz19, Meynet16} and \cite{Mink2013} for binaries. The result for our sample is shown in the Figure~\ref{f:vsini}. There is no clear dependence between magnetic field strength and rotational velocity for any type of stars. It is again one of the open questions in this field as in Section \ref{BinFrac}. We, nevertheless, can say that weakly-magnetic stars seem to rotate slower than the bulk of other stars. Mostly, this is due to the criteria by which the stars were selected for observations: in the search programs by \cite{Blazere2018}, bright stars with low v$\mathrm{\sin{i}}$ were specially selected, because shorter observations are required to obtain single spectropolarimetric image.
That is why many of these stars are Am \citep{Neiner2014, Blazere2018}. 

\subsection{Magnetic fields or magnetic fluxes?}

\begin{figure}
\center{\includegraphics[width=\columnwidth]{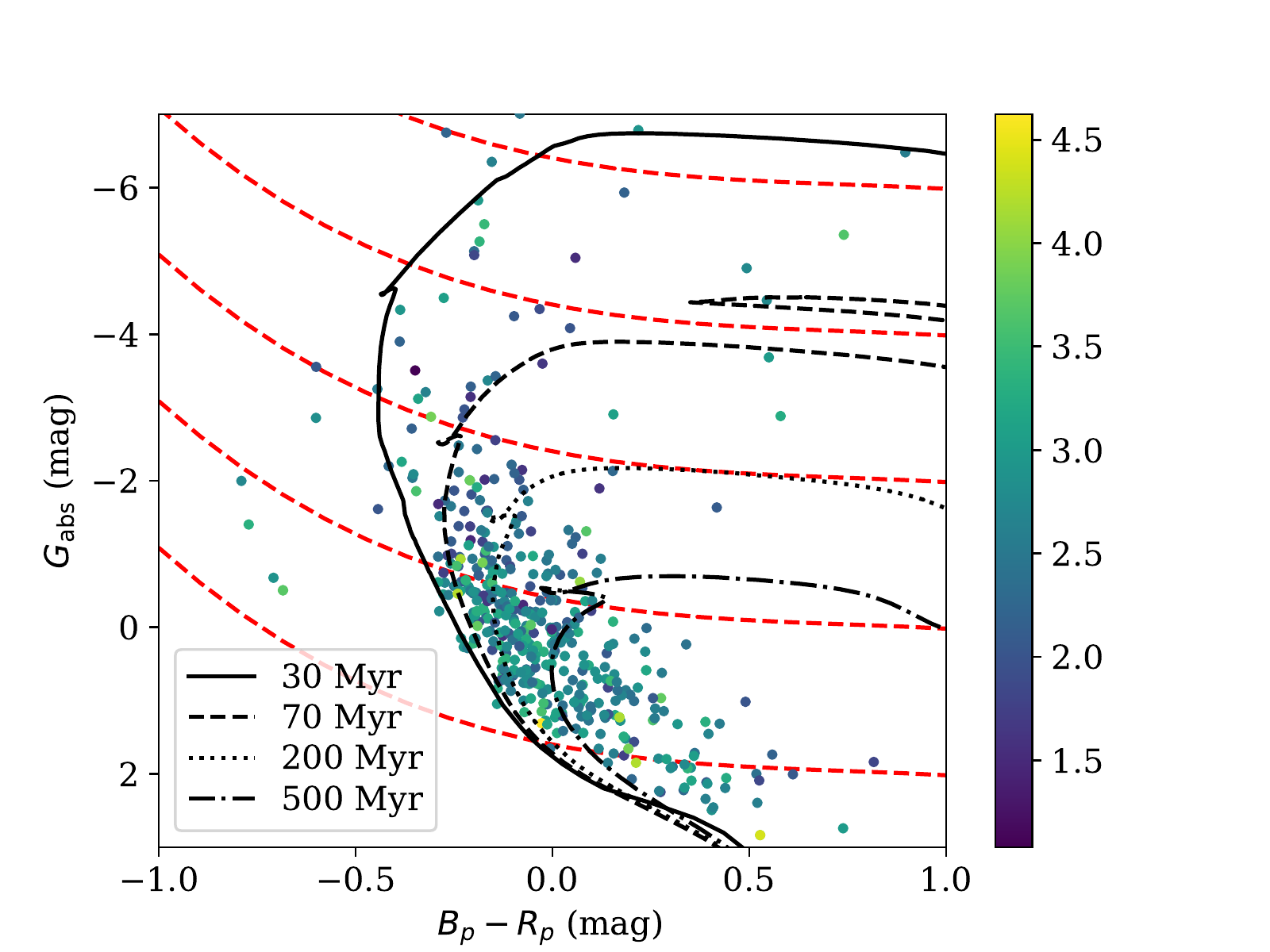}}
\caption{The Hertzsprung Russell diagram for stars with measured magnetic field. The ten-based logarithm of the rms equatorial magnetic field is shown with colour. The red dashed line corresponds to equal radii: the lower line is 1.6~$R_\odot$, next is 2.8~$R_\odot$, 6.3~$R_\odot$, 17.8~$R_\odot$, 63~$R_\odot$}
\label{f:HR}
\end{figure}

Use of the magnetic flux ($F\sim B_p R^2$) as a stellar property is a more challenging problem than analysis of the magnetic field measurements. It requires measurements of the stellar radius. Our simplified analysis shows that there is some dependence between the stellar magnetic field and the radius (the larger the radius, the weaker the magnetic field). It is important to mention here that radius of a star depends also on its mass. In fact, we might see indications of magnetic flux conservation which is studied in more detail in \cite{Landstreet2007, Fossati2016}.  We identify massive OB stars with magnetic field measurements in the Gaia DR2 and plot their absolute magnitude, colour and magnetic field in Figure~\ref{f:HR}. The radius grows approximately with the absolute magnitude while the magnetic field seems to be larger for stars with $G_\mathrm{abs}\approx 1$ than for stars with $G_\mathrm{abs}\approx -2$.

\subsection{Maximum period for magnetars}

In some models we produce magnetars with periods around $10^3$~sec, especially in the models F1-F2. Current observations seem to put an upper limit of 12 sec \citep{2013NatPh...9..431P}. However recently a magnetar type activity was discovered from 1E 161348-5055 \citep{Rea2016} with spin period 6.67 hours i.e. $\approx 24$ ksec. The formation path for this object is still unclear.
Another option could be the Poisson counting noise. If we form just two magnetars with periods $\approx 100$~sec they might not be discovered or not present in the Galactic population. 

\begin{figure}
	\includegraphics[width=\columnwidth]{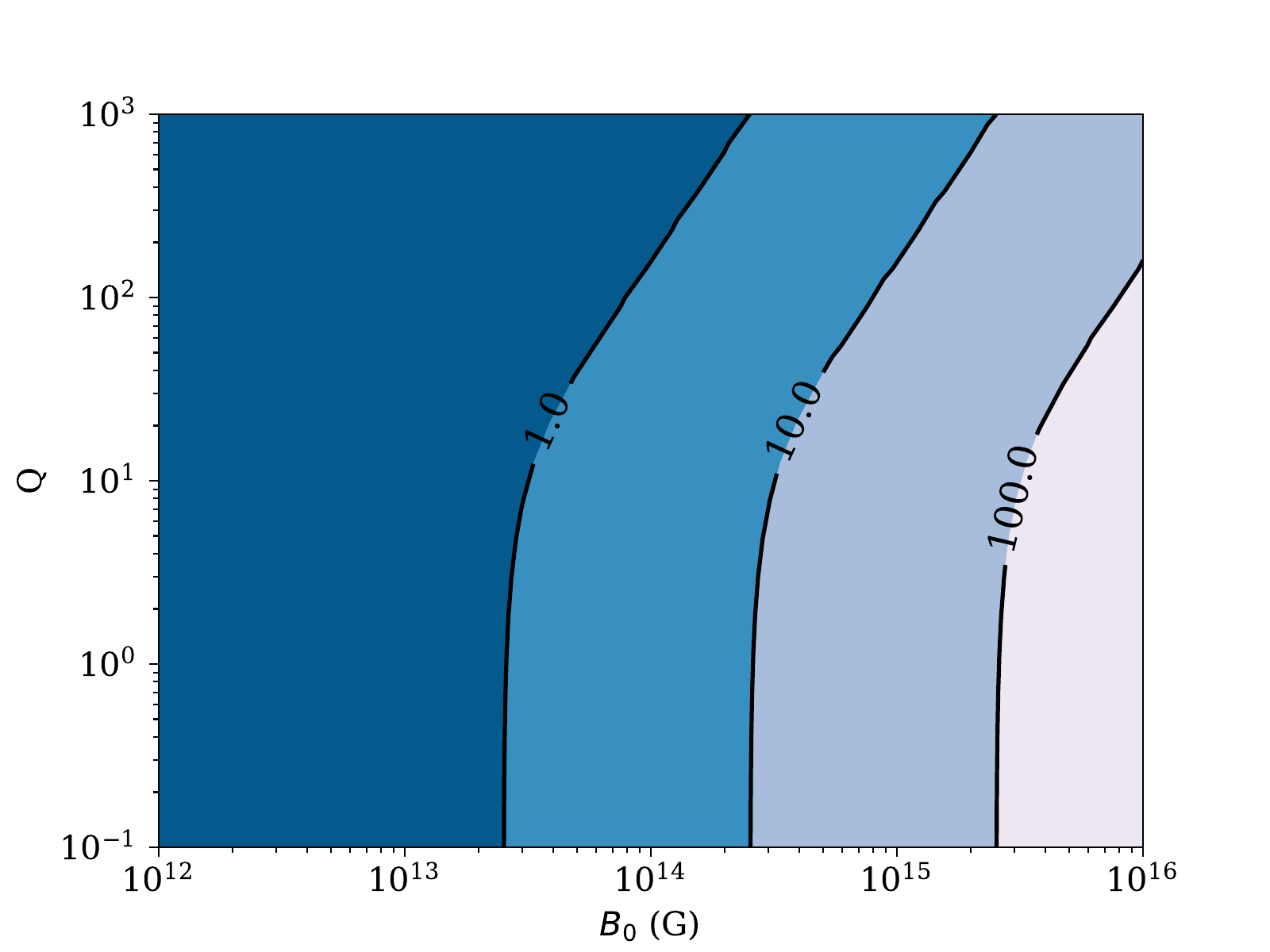}
\caption{Spin periods of NS at age 100~kyr as a function of value for initial magnetic field $B_0$ and deep crust impurity parameter $Q$. }
    \label{f:pq}
\end{figure}

On the other hand, \cite{Gullon-2015} noticed that maximum rotational periods of magnetars are related to both the initial distribution of magnetic fields and impurity parameter $Q$. Here we give a simple estimate for the maximum period of a magnetar at age $1$~Myr for fixed initial magnetic field $B_0$ and deep crust impurity parameter $Q$. We assume that the magnetic field decays simply exponentially under influence of the Ohmic term only (which is an oversimplification) due to the crust impurity according to the law:
\begin{equation}
B = B_0 \exp\left(-\frac{t}{\tau_\mathrm{Ohm}} \right).    
\end{equation}
We compute the decay timescale using eq. (\ref{e:q_decay}).
We assume that the period evolves under influence of magnetospheric torque:
\begin{equation}
P\frac{dP}{dt} = \frac{2}{3} \beta B^2.    
\label{e:p}
\end{equation}
This is a simplified expression which does not take into account the angle between orientation of the global dipole field and rotational axis of the pulsar. The exact equation is quite similar to eq.~(\ref{e:p}) and contains weak dependence on the obliquity angle. 
Combining aforementioned equations and solving the differential eq.~(\ref{e:p}) for initial period $P_0$ we get following estimate for maximum spin period:
\begin{equation}
P_\mathrm{max} = \sqrt{P_0^2 + \frac{2}{3} \tau_\mathrm{ohm} \beta B_0^2 \left(1 - \exp\left[-\frac{2t_\mathrm{max}}{\tau_\mathrm{ohm}}\right]\right)}.    
\end{equation}
We assume that $P_0$ is $0.1$~s and typical for normal radio pulsars (see e.g. \citealt{igoshev2013} for discussion of initial periods of radio pulsars). We illustrate this dependence in Figure~\ref{f:pq}. From this figure we immediately see that the initial periods can be restricted by maximum value of 12~sec if the initial distribution of magnetic fields does not include any stars with magnetic fields larger than $\approx 2\times 10^{14}$~G. Another alternative is the fast magnetic field decay: in this case the magnetic fields need to be restricted by values of $\approx 5-6\times 10^{14}$~G and $Q\approx 50$. These limits are in quantitative agreement with a cut of magnetic field distribution suggested by \cite{Gullon-2015} at $5\times 10^{14}$~G.

It is interesting to note that among known magnetars there are a few objects with estimated poloidal dipolar magnetic fields in excess of $5\times 10^{14}$~G, for example SGR 1806-20 with $B = 2\times 10^{15}$~G. If initial magnetic fields are restricted at smaller values, these cases need additional attention. During the complicated magnetic field evolution, the poloidal, dipolar component of the magnetic field can increase due to the Hall evolution, interaction with small scale fields or toroidal components. Alternatively, the vacuum dipole equation which is used to estimate the dipolar component of magnetic fields based on the period and period derivative could be less applicable to magnetars due to additional currents in magnetosphere.

\section{Conclusions}
\label{s:conclusion}

In this article, we study a hypothesis that magnetic fields of NSs have fossil origin.
In particular, we study if weakly magnetised B stars could be progenitors for normal radio pulsars and strongly magnetised B stars could be progenitors for magnetars.
To do so we collect all reliable, modern measurements of magnetic fields at surfaces on massive stars of spectral types O-A. We develop a maximum likelihood technique to estimate the parameters of the magnetic field distribution. We found that the log-normal distribution describes well measurements of magnetic fields of O and B stars. In the case of A stars, we found significant deviations from the log-normal distribution possibly related to evolution.
In the case of B stars, the parameters of the log-normal distribution are as following: $\mu_B = 2.83\pm 0.1$ $\log_{10}$~(G) i.e. $\approx 700$ and $\sigma_B = 0.65\pm 0.09$. Our results differs from \citep{Shultz2019} because we use \emph{rms} magnetic fields which are $\approx 4-5$ times smaller than polar magnetic fields.

Strongly magnetised B stars represent 5-7~per~cent of the total B population. Measurements of magnetic fields for remaining stars are challenging and results in values around a few Gauss. We collect these measurements as well and estimated the parameters of the log-normal distribution. We obtained $\mu_B = 0.14\pm 0.5$ $\log_{10}$~(G) i.e. $\approx 1.4$~G  and $\sigma_B = 0.70^{+0.57}_{-0.27}$. We notice that the difference between magnetic field of strongly magnetised B-stars and weakly magnetised B stars is 2.7 DEX. 

In accordance to the fossil field hypothesis \citep{ferrario2006} we assume that weakly-magnetised B stars produce normal radio pulsars with $\mu_B = 11.7$ and $\sigma_B=0.7$ and strongly magnetised produce magnetars with $\mu_B = 14.45$ and $\sigma_B=0.7$. To check if the resulting population looks anything like an observed population of radio pulsars and magnetars we run the population synthesis. We found that simple conservation of magnetic field in the core cannot explain the observed value of period and period derivative for normal radio pulsars. The cloud of radio pulsars is shifted towards too small period derivative values. In trying to improve our model, we guess that our original model for magnetic field conservation might be too simplistic. Therefore, we assume values of $\mu_B = 12.65$, $\sigma_B = 0.7$ for 90~per~cent of NSs and $\mu_B = 15.35$, $\sigma_B= 0.7$ for magnetars to keep 2.7~DEX difference. This model strongly overproduces bright magnetars with fluxes $S_X$ in the range $10^{-10}$~--~$10^{-8}$~erg~cm$^{-2}$~s$^{-1}$. This result does not depend on the value of the crust impurity parameter.

Therefore, we have to conclude that the fossil field hypothesis cannot simply explain NS magnetic field distribution. To correct this hypothesis it is necessary to suggest a mechanism that decreases the difference of 2.7~DEX between two groups of stars to the difference of $\approx 1$~DEX seen between magnetic fields of magnetars and normal radio pulsars e.g. \cite{Gullon-2015}. One of such mechanisms could be the field instability at the proto-NS stage. After the supernova explosion and during the first $\approx 30$~sec of its evolution the proto-NS is not transparent for neutrinos. Therefore, it cools down from the surface. The energy released inside the proto-NS is so large that convection is initiated. The convection could erase or amplify pre-existing magnetic fields. Any initial magnetic field configuration will be tested for its long-term stability at this stage. \cite{Gullon-2015} already noticed that there might exist a cut-off in a smooth distribution of magnetic fields at levels of $5\times 10^{14}$~G. A decrease in this difference could be related to a fact that magnetic fields around $10^{15}$~G are unstable at the proto-NS stage.

It is interesting to note that even if we miss most of the distribution for weakly magnetised massive stars (e.g. due to instrumental limitations) and estimate only the exponential tail, our conclusion still holds. In this case, the mean value of magnetic fields for weakly magnetised stars is located at even smaller values and the actual difference is more than 2.7 DEX

\section*{Acknowledgements}

We thank the anonymous referee for constructive comments which helped us to improve the manuscript. A.I.P. thanks the STFC for research grant ST/S000275/1. Authors are grateful to Hagai Perets and Sergei Popov for multiple insightful discussions. A.I.P. and E.I.M. thank the Technion - Israel Institute of Technology for hosting them during a part of this research.
A.F.K. thanks the RFBR grant 19-02-00311~A for the support. E.I.M. acknowledge funding by the European Research Council through ERC Starting Grant No. 679852 'RADFEEDBACK'. E.I.M. would like to thank D. Sz{\'e}csi and M. Shultz for the productive conversations about evolution of massive stars. E.I.M. and A.P.I would also like to thank Dr B. Gaches and A. Frantsuzova for comments which helped to improve this manuscript.
This work has made use of data from the European Space Agency (ESA) mission
{\it Gaia} (\url{https://www.cosmos.esa.int/gaia}), processed by the {\it Gaia}
Data Processing and Analysis Consortium (DPAC,
\url{https://www.cosmos.esa.int/web/gaia/dpac/consortium}). Funding for the DPAC
has been provided by national institutions, in particular the institutions
participating in the {\it Gaia} Multilateral Agreement. 

\section{Data availability}
The data underlying this article are available in the article and in its online supplementary material. The code NINA for population synthesis of isolated radio pulsars is available on Github. Code for population synthesis of magnetars is available upon reasonable request.

\bibliographystyle{mnras} 
\bibliography{references.bib}



\appendix

\section{Statistical tools}
\label{s:statistics}
\begin{figure}
	\includegraphics[width=\columnwidth]{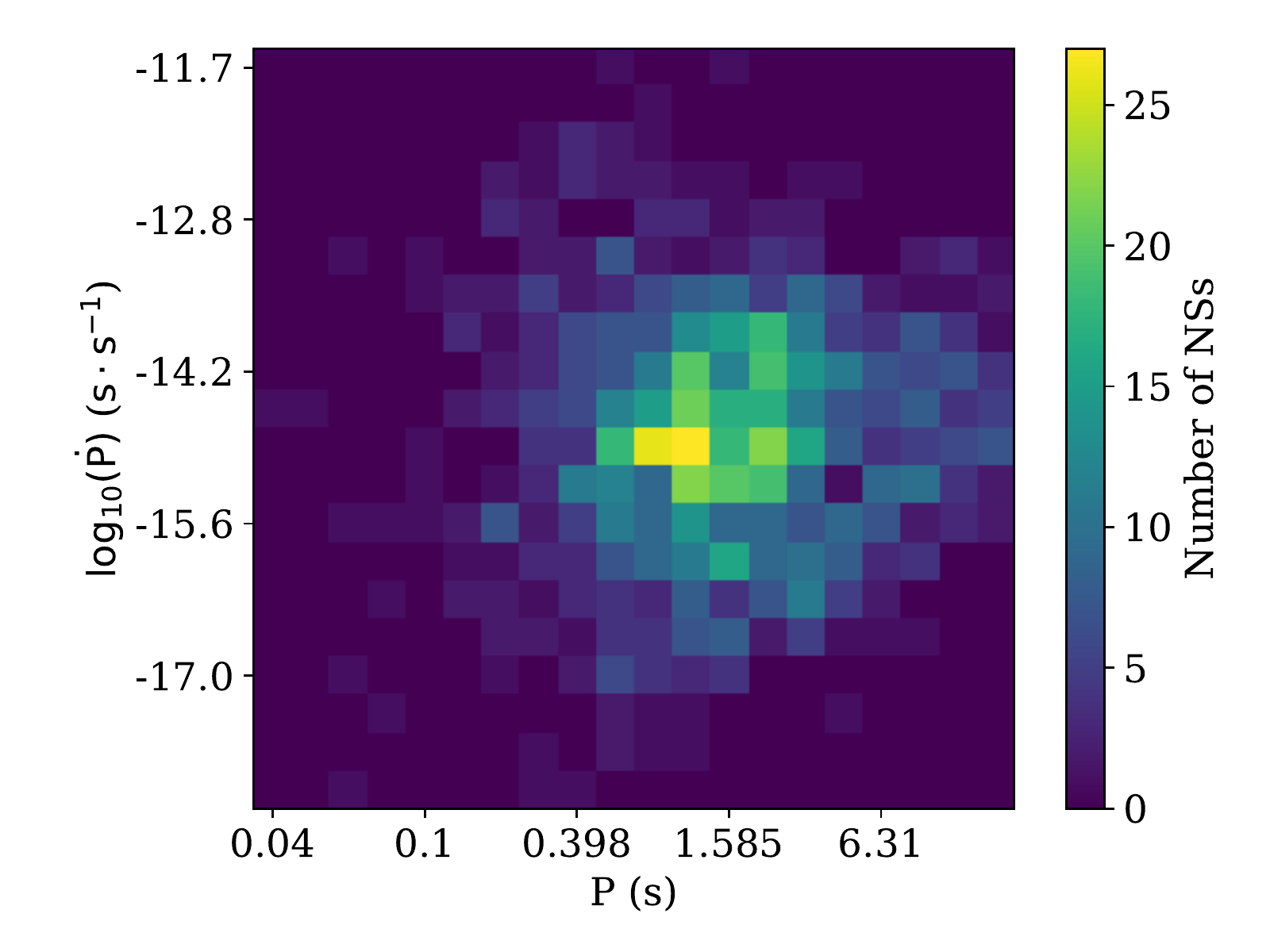}
\caption{The distribution of pulsars detected in Swinburn and Parkes radio surveys over two-dimensional bins (number of pulsars per bin is shown with color) in logarithm of period and period derivative.}
    \label{f:bins}
\end{figure}

A comparison of synthetic radio pulsar and magnetar populations with actual data is challenging in a few aspects. Two fundamental measurements ($P$ and $\dot P$) have negligible measurement uncertainties, so the actual $P$~--~$\dot P$ diagram is defined by age of individual objects and selectional effects rather than measurement uncertainties. The shape of the distribution is strongly affected by a choice of initial periods and magnetic field distributions, see e.g. \cite{Gullon-2015}. At the moment, with $\approx 2500$ known objects from total population of radio pulsars, we are often in regime where only a few pulsars are detected in a particular region of $P$~~-~$\dot P$ i.e. our picture is affected by shot (Poisson) noise in details. In earlier population synthesis \citep{Gullon-2015, 2006ApJ...643..332F} researcher used either two-dimensional Kolmogorov-Smirnov test, or compared separately distributions in period and period derivative. Often the Kolmogorov Smirnov test was used as a criterion to optimise the parameters of the population synthesis instead of analysis of individual hypothesis. In our population synthesis we want to take the Poisson noise seriously and use C-stat. To do so, we divide the $P$~--~$\dot P$ diagram into two-dimensional bins (see Fig.~\ref{f:bins}) and compute following:
\begin{equation}
C = 2\sum_{i=1}^{M} N_\mathrm{synth} - N_\mathrm{obs} \log (N_\mathrm{synth}) + \log (N_\mathrm{obs}!).
\end{equation}
where $M$ is number of bins, $N_\mathrm{synth}$ -- number of synthetic radio pulsars in the bin and $N_\mathrm{obs}$ actual number of observed radio pulsars. 
If in a particular bin there is no synthetic and observed pulsars, we assume that this bin do not contribute to the statistics. If a synthetic population do not predict any pulsar in a particular bin, but we do observe one or more pulsars there, we assume that theoretical probability to find a pulsar in this bin is $1/M$. This choice is arbitrary and it is one of the reasons why we compute the size of confidential intervals using additional simulations.

To test which values of C statistics are acceptable and which are not we perform a bootstrapping. We compare the value of C statistics found when only Parkes and Swinburne survey are analysed and when the whole ATNF catalogue is analysed. This comparison gives the value $C = 4.13$ .

\section{Corrected radiative transfer formulas}\label{A}
The correct form of the transmitted flux $n_+$ and the reflected flux $n_-$ from the formula (36) \cite{Lyutikov}:
\begin{equation}
\begin{gathered}
n_+ (\omega, \omega_0) d\omega = e^{- \tau_0/2} \left( \delta(\omega -\omega_0) + \frac{\tau_0}{8 \beta_\mathrm{T}} \sqrt{\frac{\omega_0 (1 + 4\beta_\mathrm{T})- \omega}{\omega-\omega_0}} \cdot \right. \\
\left. I_1 \left( {\frac{\tau_0}{4 \beta_\mathrm{T}\omega_0}} \sqrt{(\omega-\omega_0) \cdot (\omega_0(1+ 4\beta_\mathrm{T})- \omega)} \right) \right) \frac{d\omega}{\omega_0},
\end{gathered}
\end{equation}

$$
n_- (\omega, \omega_0) d\omega = \frac{\tau_0}{8 \beta_\mathrm{T}} e^{- \tau_0/2} I_0 \left( {\frac{\tau_0}{4 \beta_\mathrm{T}\omega_0}} \sqrt{(\omega_0(1+ 2\beta_\mathrm{T})- \omega)} \right. \hspace{4cm}
$$
\begin{equation}
 \left. \sqrt{ (\omega-\omega_0(1-2\beta_\mathrm{T}))} \right) \frac{d\omega}{\omega_0},
\end{equation}
where $I_0$ and $I_1$ are the modified Bessel functions.

\section{Catalogue of stars with measurements of magnetic fields}
\label{s:catalogue}

\begin{table*}
\caption{List of B stars} \label{tab:B_stars}
\begin{tabular}{lccccccc}
    \hline
    Name & $B_{\mathrm{rms}}$ $\pm$ $\sigma_B$ & P$_{\mathrm{orbital}}$ & Ra & Dec & $\mathrm{v \sin{i} \pm \sigma_v}$  & Ref. \\
    &  [G] & [days] & & & [km/s] & \\
    \hline\\

    HD147933  &  480$\pm$80  &  876000  &  16h25m35s  &  -23d26m49s  &  196$\pm$10  &    \cite{Brown_v} \\
    &&&&&& \cite{Hubrig2018} \\

    HD3360  &  20$\pm$4  &  -  &  0h36m58s  &  53d53m48s  &  19$\pm$2  &    \cite{Shultz18} \\

    HD23478  &  2100$\pm$200  &  -  &  3h46m40s  &  32d17m24s  &  136$\pm$7  &    \cite{Shultz18} \\

    HD25558  &  100$\pm$40  &  3285  &  4h3m44s  &  5d26m8s  &  35$\pm$4  &    \cite{Shultz18} \\

    HD35298  &  3200$\pm$300  &  -  &  5h23m50s  &  2d4m55s  &  58$\pm$2  &    \cite{Shultz18} \\

    HD35502  &  1580$\pm$90  &  14600  &  5h25m1s  &  -2d48m55s  &  78$\pm$5  &    \cite{Shultz18} \\

    HD35912  &  500$\pm$200  &  -  &  5h28m1s  &  1d17m53s  &  15$\pm$1  &    \cite{Nieva_v} \\
    &&&&&& \cite{Shultz18}\\

    HD36485  &  2570$\pm$70  &  25.592  &  5h32m0s  &  -0d17m4s  &  26$\pm$4  &    \cite{Shultz18}\\

    HD36526  &  3400$\pm$200  &  -  &  5h32m13s  &  -1d36m1s  &  55$\pm$5  &    \cite{Shultz18} \\

    HD36982  &  340$\pm$80  &  -  &  5h35m9s  &  -5d27m53s  &  86$\pm$5  &    \cite{Shultz18} \\

    HD37017  &  1800$\pm$300  &  18.6556  &  5h35m21s  &  -4d29m39s  &  134$\pm$15  &    \cite{Shultz18} \\

    HD37058  &  750$\pm$50  &  -  &  5h35m33s  &  -4d50m15s  &  11$\pm$2  &    \cite{Shultz18} \\

    HD37061  &  500$\pm$80  &  19.139  &  5h35m31s  &  -5d16m2s  &  189$\pm$8  &  \cite{Shultz18}\\

    HD37479  &  2760$\pm$90  &  -  &  5h38m47s  &  -2d35m40s  &  145$\pm$5  &    \cite{Shultz18} \\

    HD37776  &  1700$\pm$200  &  -  &  5h40m56s  &  -1d30m25s  &  91$\pm$4  &    \cite{Shultz18} \\

    HD44743  &  26$\pm$3  &  -  &  6h22m41s  &  -17d57m21s  &  20$\pm$7  &    \cite{Shultz18} \\

    HD46328  &  350$\pm$10  &  -  &  6h31m51s  &  -23d25m6s  &  8  &    \cite{Shultz18} \\


    HD55522  &  900$\pm$100  &  -  &  7h12m12s  &  -25d56m33s  &  70$\pm$2  &    \cite{Shultz18} \\

    HD58260  &  1820$\pm$30  &  -  &  7h23m19s  &  -36d20m24s  &  3$\pm$2  &    \cite{Shultz18} \\

    HD61556  &  820$\pm$30  &  -  &  7h38m49s  &  -26d48m13s  &  58$\pm$3  &    \cite{Shultz18} \\

    HD63425  &  130$\pm$6  &  -  &  7h47m7s  &  -41d30m13s  &  3$\pm$3  &    \cite{Shultz18}\\

    HD64740  &  820$\pm$90  &  -  &  7h53m3s  &  -49d36m46s  &  135$\pm$2  &    \cite{Shultz18} \\

    HD66522  &  620$\pm$70  &  -  &  8h1m35s  &  -50d36m20s  &  3$\pm$2  &    \cite{Shultz18}\\

    HD66665  &  126$\pm$9  &  -  &  8h4m47s  &  6d11m9s  &  8$\pm$2  &    \cite{Shultz18} \\

    HD66765  &  810$\pm$30  &  ?  &  8h2m55s  &  -48d19m29s  &  95$\pm$3  &    \cite{Shultz18} \\
    &&&&&& \cite{Alecian14}\\

    HD67621  &  290$\pm$30  &  -  &  8h6m41s  &  -48d29m50s  &  21$\pm$5  &    \cite{Shultz18}\\
    &&&&&& \cite{Alecian14}\\

    HD96446  &  1000$\pm$20  &  -  &  11h6m5s  &  -59d56m59s  &  6$\pm$2  &    \cite{HD96446_v} \\
    &&&&&& \cite{Shultz18}\\

    HD105382  &  740$\pm$30  &  -  &  12h8m5s  &  -50d39m40s  &  74$\pm$5  &    \cite{Shultz18} \\

    HD121743  &  330$\pm$90  &  -  &  13h58m16s  &  -42d6m2s  &  74$\pm$4  &    \cite{Shultz18} \\
    &&&&&& \cite{Alecian14}\\

    HD122451  &  30$\pm$10  &  357.02  &  14h3m49s  &  -60d22m22s  &  70$\pm$10  &    \cite{Shultz18} \\

    HD125823  &  530$\pm$70  &  -  &  14h23m2s  &  -39d30m42s  &  16$\pm$2  &    \cite{Brown_v} \\
    &&&&&& \cite{Shultz18}\\

    HD127381  &  170$\pm$50  &  -  &  14h32m37s  &  -50d27m25s  &  62$\pm$4  &     \cite{Shultz18} \\

    HD130807  &  1000$\pm$30  &  ?  &  14h51m38s  &  -43d34m31s  &  27$\pm$3  &    \cite{Brown_v} \\
    &&&&&& \cite{Shultz18}\\

    HD136504A  &  230$\pm$20  &  4.56  &  15h22m40s  &  -44d41m22s  &  41$\pm$6  &    \cite{Brown_v} \\
    &&&&&& \cite{Shultz18}\\

    HD136504B  &  140$\pm$50  &  4.56  &  15h22m40s  &  -44d41m22s  &  177$\pm$17  &    \cite{Brown_v} \\
    &&&&&& \cite{Shultz18}\\

    HD142184  &  2010$\pm$100  &  -  &  15h53m55s  &  -23d58m41s  &  288$\pm$6  &    \cite{Shultz18} \\

    HD142990  &  1200$\pm$100  &  -  &  15h58m34s  &  -24d49m53s  &  122$\pm$2  &    \cite{Shultz18} \\

    HD149277  &  2900$\pm$200  &  25.390  &  16h35m48s  &  -45d40m43s  &  8$\pm$3  &   \cite{Shultz18} \\

    HD149438  &  88$\pm$4  &  -  &  16h35m52s  &  -28d12m57s  &  7$\pm$3  &    \cite{Shultz18} \\

    HD156324  &  2700$\pm$400  &  -  &  17h18m23s  &  -32d24m14s  &  50$\pm$10  &    \cite{Shultz18} \\
    
    HD156424A  &  1550$\pm$219  &  ?  &  17h18m54s  &  -32d20m44s  &  5$\pm$1 & \cite{shultz2020mobster} \\

    HD156424B  &  26$\pm$36  &  ?  &  17h18m54s  &  -32d20m44s  &  25$\pm$5  & \cite{shultz2020mobster} \\

    HD163472  &  300$\pm$100  &  -  &  17h56m18s  &  0d40m13s  &  62$\pm$5  &    \cite{Shultz18} \\

    HD164492C  &  1800$\pm$100  &  12.5351  &  18h2m23s  &  -23d1m50s  &  138$\pm$10  &    \cite{Shultz18} \\

    HD175362  &  5130$\pm$30  &  -  &  18h56m40s  &  -37d20m35s  &  34$\pm$4  &    \cite{Shultz18} \\

    HD176582  &  1640$\pm$60  &  -  &  18h59m12s  &  39d13m2s  &  103$\pm$7  &    \cite{Shultz18} \\
    & & & & & &  \cite{Bohlender11} \\

    HD182180  &  2700$\pm$100  &  -  &  19h24m30s  &  -27d51m57s  &  306$\pm$5  &    \cite{Shultz18} \\

    HD184927  &  1820$\pm$70  &  -  &  19h35m32s  &  31d16m35s  &  8$\pm$2  &    \cite{Shultz18} \\

    HD186205  &  820$\pm$30  &  -  &  19h42m37s  &  9d13m39s  &  4$\pm$4  &    \cite{Shultz18} \\

    HD189775  &  1290$\pm$70  &  -  &  19h59m15s  &  52d3m20s  &  58$\pm$2  &    \cite{Shultz18} \\

    HD205021  &  76$\pm$6  &  33580  &  21h28m39s  &  70d33m38s  &  37$\pm$2  &    \cite{Shultz18}\\
    \hline
 \end{tabular}
\end{table*}

\begin{table*}
\contcaption{List of B stars} \label{tab:B_stars}
\begin{tabular}{lccccccc}
    \hline
    Name & $B_{\mathrm{rms}}$ $\pm$ $\sigma_B$ & P$_{\mathrm{orbital}}$ & Ra & Dec & $\mathrm{v \sin{i} \pm \sigma_v}$  & Ref. \\
    &  [G] & [days] & & & [km/s] & \\
    \hline\\

    HD208057  &  130$\pm$30  &  -  &  21h53m3s  &  25d55m30s  &  105$\pm$5  &    \cite{Shultz18} \\

    ALS3694  &  3700$\pm$1000  &  -  &  16h40m33s  &  -48d53m16s  &  48$\pm$3  &    \cite{Shultz18} \\

    HD34736  &  4700$\pm$350  &  -  &  5h19m21s  &  -7d20m50s  &  73$\pm$7  &  \cite{Romanyuk18} \\

    HD35456  &  441$\pm$96  &  ?  &  5h24m40s  &  -2d29m52s  &  25$\pm$2  &    \cite{Semenko_v} \\
    &&&&&& \cite{Romanyuk18}\\

    HD36313  &  1020$\pm$450  &  -  &  5h30m45s  &  0d22m24s  &  170$\pm$30  &    \cite{Romanyuk_v} \\
    &&&&&& \cite{Romanyuk18}\\

    HD36997  &  1227$\pm$87  &  ?  &  5h35m13s  &  -2d22m52s  &  -  &  \cite{Romanyuk18} \\

    HD37687  &  560$\pm$35  &  -  &  5h40m19s  &  -3d25m37s  &  25$\pm$5  &    \cite{Romanyuk_v} \\
    &&&&&& \cite{Romanyuk18}\\

    HD40759  &  1990$\pm$240  &  -  &  6h0m45s  &  -3d53m44s  &  -  & \cite{Romanyuk18} \\

    HD290665  &  1700$\pm$100  &  -  &  5h35m10s  &  -0d50m12s  &  -  &  \cite{Romanyuk18} \\

    HD14437  &  6610$\pm$350  &  -  &  2h21m2s  &  42d56m38s  &  5.1  &  \cite{Cho19} \\

    HD279277  &  8100$\pm$620  &  -  &  4h2m13s  &  37d9m57s  &  -  &    \cite{Cho19} \\

    HD282151  &  5410$\pm$410  &  -  &  4h29m41s  &  30d41m2s  &  -  &   \cite{Cho19} \\

    HD35379  &  5710$\pm$910  &  -  &  5h25m38s  &  30d41m14s  &  -  &   \cite{Cho19} \\

    HD252382  &  5600$\pm$570  &  -  &  6h9m12s  &  20d26m52s  &  -  &    \cite{Cho19} \\

    HD263064  &  5390$\pm$510  &  -  &  6h44m45s  &  11d35m34s  &  -  &   \cite{Cho19} \\

    HD235936  &  5090$\pm$1170  &  -  &  22h43m1s  &  51d41m12s  &  -  &   \cite{Cho19} \\

    HD216001  &  6990$\pm$510  &  -  &  22h48m6s  &  59d6m42s  &  -  &    \cite{Cho19} \\

    HD201250  &  7810$\pm$690  &  -  &  21h6m36s  &  48d37m40s  &  -  &    \cite{Cho19} \\

    HD350689  &  5150$\pm$480  &  -  &  19h51m9s  &  18d27m10s  &  -  &   \cite{Cho19} \\

    HD92938  &  117$\pm$48  &  -  &  10h42m14s  &  -64d27m59s  &  120  &    \cite{Zboril_v} \\
    &&&&&& \cite{Hubrig2014}\\

    HD64503  &  127$\pm$38  &  ?  &  7h52m38s  &  -38d51m46s  &  187 &  \cite{Hubrig2014} \\

    HD58647  &  212$\pm$64  &  -  &  7h25m56s  &  -14d10m43s  &  - &   \cite{Hubrig2014} \\

    HD133518  &  476$\pm$13  &  -  &  15h6m55s  &  -52d1m47s  &  - &   \cite{Alecian14} \\

    HD147932  &  925$\pm$215  &  -  &  16h25m35s  &  -23d24m18s  &  180 &  \cite{Alecian14} \\

    HD32633  &  2159$\pm$26  &  -  &  5h6m8s  &  33d55m7s  &  -  &   \cite{Cho19} \\

    HD318100  &  580.5$\pm$59  &  -  &  17h40m12s  &  -32d9m32s  &  42 &  \cite{landstreet08} \\

    HD162576  &  12.5$\pm$25  &  -  &  17h53m15s  &  -34d37m15s  &  - &   \cite{landstreet08} \\

    HD35502  &  1793$\pm$458  &  14600  &  5h25m1s  &  -2d48m55s  &  12$\pm$2  &    \cite{Sikora16} \\

    OGLE\_SMC  &  1350$\pm$595  &  ?  &  0h54m2s  &  -72d42m22s  &  - &  \cite{Bagnulo17}   \\
    SC6 311225 & & & & & & \\

    MACHO &  605$\pm$325  &  -  &  5h13m26s  &  -69d21m55s  &  - &  \cite{Bagnulo17}  \\
    5.5377.4508  & & & & & & \\

    HD18296  &  213$\pm$20  &  -  &  2h57m17s  &  31d56m3s  &  25  &    \cite{Helmut_v} \\
    & & & & & &\cite{Auriere_v} \\

    HD39317  &  216$\pm$59  &  -  &  5h52m22s  &  14d10m18s  &  45$\pm$2  &    \cite{Auriere_v} \\

    HD43819  &  628$\pm$25  &  -  &  6h19m1s  &  17d19m30s  &  10$\pm$2  &  \cite{Auriere_v}  \\

    HD68351  &  325$\pm$47  &  -  &  8h13m8s  &  29d39m23s  &  33$\pm$2  &  \cite{Auriere_v}  \\

    HD148112  &  204$\pm$21  &  -  &  16h25m24s  &  14d1m59s  &  44.5$\pm$1  &    \cite{Auriere_v} \\

    HD179527  &  156$\pm$46  &  -  &  19h11m46s  &  31d17m0s  &  33$\pm$2  &    \cite{Auriere_v} \\

    HD183056  &  290$\pm$42  &  -  &  19h26m9s  &  36d19m4s  &  26$\pm$2  &    \cite{Auriere_v} \\

    HD35411A  &  52.6$\pm$32.7  &  2687.3  &  5h24m28s  &  -2d23m49s  &  35  &    \cite{Helmut_v} \\

    HD32650  &  91$\pm$18  &  -  &  5h12m22s  &  73d56m48s  &  30$\pm$2  &  \cite{Auriere_v} \\

    HD38170  &  74$\pm$20  &  1.38  &  16h5m49s  &  0d34m23s  &  65$\pm$9  &    \cite{Kobzar_v}\\
    & & & & & &\cite{Davia-Uraz2020} \\
    
    HD 62658A & 136$\pm$40 & 4.75 & - & - & 26.2$\pm$1.3 & \cite{Shultz2019} \\
    \hline
 \end{tabular}
\end{table*}

\begin{table*}
\caption{List of A stars} \label{tab:A_stars}
\begin{tabular}{lcccccc}
    \hline
    Name & $B_{\mathrm{rms}}$ $\pm$ $\sigma_B$ & P$_{\mathrm{orbital}}$ & Ra & Dec & Ref. \\
    &  [G] & [days] & & &  \\
    \hline\\
    HD15089  &  109$\pm$63  &  -  &  2h29m3s  &  67d24m8s  &  \cite{Sikora19}\\

	HD15144  &  581.6$\pm$7.2  &  -  &  2h26m0s  &  -15d20m28s  & \cite{Sikora19}  \\

	HD24712  &  560$\pm$160  &  -  &  3h55m16s  &  -12d5m56s  &  \cite{Rusomarov13} \\
	&&&&& \cite{Sikora19} \\

	HD56022  &  79$\pm$39  &  -  &  7h13m13s  &  -45d10m57s  & \cite{Sikora19} \\

	HD72968  &  427$\pm$16  &  -  &  8h35m28s  &  -7d58m56s  &  \cite{Auriere_v}\\

	HD96616  &  79$\pm$18  &  -  &  11h7m16s  &  -42d38m19s  & \cite{Sikora19} \\

	HD27309  &  804$\pm$50  &  1.5689  &  4h19m36s  &  21d46m24s  & \cite{Auriere_v} \\

	HD112413  &  104$\pm$96  &  -  &  12h56m1s  &  38d19m6s  & \cite{Sikora19} \\

	HD118022  &  533$\pm$55  &  -  &  13h34m7s  &  3d39m32s  & \cite{Sikora19} \\

	HD119213  &  380$\pm$110  &  -  &  13h40m21s  &  57d12m27s  &  \cite{Sikora19} \\

	HD130559  &  280$\pm$25  &  25.3992  &  14h49m19s  &  -14d8m56s  & \cite{Sikora19} \\

	HD137949  &  1620$\pm$100  &  4.8511  &  15h29m34s  &  -17d26m27s  & \cite{Sikora19} \\

	HD152107  &  961$\pm$49  &  -  &  16h49m14s  &  45d58m59s  & \cite{Sikora19} \\

	HD223640  &  420$\pm$350  &  -  &  23h51m21s  &  -18d54m32s  &  \cite{Sikora19} \\

	HD2453  &  860$\pm$52  &  $>$ 2453  &  0h28m28s  &  32d26m15s  & \cite{Mathys17}  \\

	HD12288  &  1643$\pm$120  &  1546.5  &  2h3m30s  &  69d34m56s  & \cite{Mathys17} \\

	HD116458  &  1552$\pm$66  &  126.233  &  13h25m50s  &  -70d37m38s  & \cite{Mathys17}  \\

	HD142070  &  376$\pm$73  &  $>$ 2500  &  15h52m35s  &  -1d1m52s  & \cite{Mathys17} \\

	HD47103  &  2778$\pm$422  &  -  &  6h37m44s  &  19d56m55s  & \cite{Mathys17} \\

	HD50169  &  754$\pm$69  &  1764  &  6h51m59s  &  -1d38m40s  & \cite{Mathys17} \\

	HD55719  &  725$\pm$87  &  46.31803  &  7h12m15s  &  -40d29m55s  & \cite{Mathys17} \\

	HD61468  &  1884$\pm$136  &  27.2728  &  7h38m22s  &  -27d52m7s  & \cite{Mathys17} \\

	HD75445  &  135$\pm$57  &  -  &  8h48m42s  &  -39d14m1s  & \cite{Mathys17} \\

	HD81009  &  1870.5$\pm$193.5  &  10700  &  9h22m50s  &  -9d50m19s  & \cite{Mathys17}  \\

	HD93507  &  2227$\pm$209.5  &  -  &  10h45m50s  &  -68d7m49s  & \cite{Mathys17} \\

	HD94660  &  1864$\pm$83  &  848.96  &  10h55m1s  &  -42d15m4s  & \cite{Mathys17} \\

	HD110066  &  117$\pm$62  &  -  &  12h39m16s  &  35d57m7s  & \cite{Mathys17} \\

	HD119027  &  510$\pm$55  &  -  &  13h41m19s  &  -28d46m59s  & \cite{Mathys17} \\

	HD126515  &  1660$\pm$122.5  &  -  &  14h25m55s  &  0d59m33s  & \cite{Mathys17} \\

	HD187474  &  1589$\pm$84  &  689.68  &  19h51m50s  &  -39d52m27s  & \cite{Mathys17} \\

	HD192678  &  1451$\pm$80  &  -  &  20h13m36s  &  53d39m33s  & \cite{Mathys17} \\

	HD225114  &  7430$\pm$320  &  -  &  0h3m38s  &  70d18m21s  & \cite{Cho19} \\

	HD2887  &  4950$\pm$240  &  ?  &  0h32m33s  &  55d12m53s  & \cite{Cho19} \\

	HD8700  &  8690$\pm$330  &  -  &  1h22m59s  &  -73d35m2s  & \cite{Cho19} \\

	HD14873  &  5630$\pm$350  &  -  &  2h25m29s  &  47d59m11s  & \cite{Cho19} \\

	HD18410  &  5010$\pm$410  &  -  &  2h59m45s  &  54d19m44s  & \cite{Cho19} \\

	HD25092  &  6090$\pm$400  &  -  &  4h1m16s  &  46d55m13s  & \cite{Cho19} \\

	HD25706  &  9330$\pm$500  &  -  &  4h6m42s  &  45d46m40s  & \cite{Cho19} \\

	HD27404  &  5290$\pm$460  &  -  &  4h20m37s  &  28d53m31s  & \cite{Cho19} \\

	HD31552  &  4160$\pm$440  &  -  &  4h58m1s  &  30d7m32s  & \cite{Cho19} \\

	HD31629  &  7920$\pm$650  &  -  &  5h1m32s  &  63d1m16s  & \cite{Cho19} \\

	HD241957  &  5000$\pm$460  &  -  &  5h14m54s  &  16d5m11s  & \cite{Cho19} \\

	HD241843  &  3900$\pm$330  &  -  &  5h15m10s  &  33d0m47s  & \cite{Cho19} \\

	HD243096  &  5800$\pm$650  &  -  &  5h22m54s  &  13d46m38s  & \cite{Cho19} \\

	HD41613  &  5410$\pm$370  &  -  &  5h55m50s  &  -77d50m43s  & \cite{Cho19} \\

	HD291513  &  5250$\pm$480  &  -  &  6h17m17s  &  -2d14m31s  & \cite{Cho19} \\

	HD47774  &  8780$\pm$1360  &  -  &  6h41m12s  &  24d5m3s  & \cite{Cho19} \\

	HD50169  &  4400$\pm$260  &  -  &  6h51m59s  &  -1d38m40s  & \cite{Cho19} \\

	HD266311  &  4430$\pm$250  &  -  &  6h54m58s  &  4d8m27s  & \cite{Cho19} \\

	HD318820  &  5050$\pm$620  &  -  &  18h1m6s  &  -31d37m30s  & \cite{Cho19} \\

	HD4778  &  1137$\pm$20  &  -  &  0h50m18s  &  45d0m8s  & \cite{Silvester12} \\

	HD40312  &  177.5$\pm$11.5  &  -  &  5h59m43s  &  37d12m45s  & \cite{Kochukhov19} \\

	HD62140  &  1031$\pm$19  &  -  &  7h46m27s  &  62d49m49s  & \cite{Sikora19} \\

	HD71866  &  1249$\pm$22  &  -  &  8h31m10s  &  40d13m29s  & \cite{Silvester12} \\

	HD33629  &  4760$\pm$200  &  -  &  5h10m5s  &  -33d46m45s  & \cite{F2008} \\

	HD42075  &  8540$\pm$20  &  -  &  6h7m36s  &  -26d37m15s  & \cite{F2008} \\

	HD44226  &  4990$\pm$150  &  -  &  6h19m34s  &  -25d19m42s  & \cite{F2008} \\

	HD46665  &  4630$\pm$130  &  -  &  6h33m41s  &  -22d41m45s  & \cite{F2008} \\
	\hline
 \end{tabular}
\end{table*}

\begin{table*}
\contcaption{List of A stars} \label{tab:A_stars}
\begin{tabular}{lcccccc}
    \hline
    Name & $B_{\mathrm{rms}}$ $\pm$ $\sigma_B$ & P$_{\mathrm{orbital}}$ & Ra & Dec  & Ref. \\
    &  [G] & [days] & & & \\
    \hline\\
	HD47009  &  7360$\pm$150  &  ?  &  6h35m48s  &  -13d44m49s  & \cite{F2008} \\

	HD52847  &  4440$\pm$10  &  -  &  7h1m46s  &  -23d6m20s  & \cite{F2008} \\

	HD55540  &  12730$\pm$300  &  -  &  7h12m30s  &  -21d3m53s  & \cite{F2008} \\

	HD69013  &  4800$\pm$100  &  -  &  8h14m28s  &  -15d46m31s  & \cite{F2008} \\

	HD72316  &  5180$\pm$400  &  -  &  8h30m58s  &  -33d38m3s  & \cite{F2008} \\

	HD88701  &  4380$\pm$350  &  -  &  10h13m0s  &  -37d30m12s  & \cite{F2008} \\

	HD110274  &  4020$\pm$380  &  -  &  12h41m30s  &  -58d55m24s  & \cite{F2008} \\

	HD117290  &  6380$\pm$20  &  -  &  13h30m13s  &  -49d7m58s  & \cite{F2008} \\

	HD121661  &  6160$\pm$1140  &  -  &  13h58m42s  &  -62d43m7s  & \cite{F2008} \\

	HD135728B  &  3630$\pm$300  &  ?  &  15h17m38s  &  -31d27m32s  & \cite{F2008} \\

	HD143487  &  4230$\pm$70  &  ?  &  16h1m44s  &  -30d54m56s  & \cite{F2008} \\

	HD218994A  &  440$\pm$23  &  ?  &  23h13m16s  &  -60d35m2s  &  \cite{Kurtz2008}\\

	HD16605  &  2152$\pm$32  &  -  &  2h40m58s  &  42d52m16s  &  \cite{landstreet08} \\

	HD108945  &  100$\pm$26  &  -  &  12h31m0s  &  24d34m1s  & \cite{Sikora19} \\
	&&&&& \cite{landstreet08} \\

	HD153948  &  202.5$\pm$87  &  -  &  17h4m15s  &  -38d3m6s  & \cite{landstreet08}  \\

	HD317857  &  919$\pm$20  &  -  &  17h34m34s  &  -32d36m8s  &  \cite{landstreet08} \\

	HD162725  &  61$\pm$17  &  -  &  17h53m58s  &  -34d49m51s  & \cite{landstreet08}  \\

	HD169842  &  190$\pm$24  &  -  &  18h26m22s  &  6d51m25s  & \cite{landstreet08} \\

	HD169959A  &  691$\pm$113  &  ?  &  18h26m52s  &  6d25m24s  & \cite{landstreet08}  \\

	HD94660  &  1890$\pm$21  &  -  &  10h55m1s  &  -42d15m4s  &  \cite{Bagnulo17} \\

	HD8441  &  157$\pm$18  &  -  &  1h24m18s  &  43d8m31s  &  \cite{Auriere_v}\\

	HD10221  &  148$\pm$34  &  -  &  1h42m20s  &  68d2m34s  & \cite{Auriere_v} \\

	HD22374  &  523$\pm$24  &  -  &  3h36m58s  &  23d12m39s  &  \cite{Auriere_v} \\

	HD32549  &  186$\pm$39  &  -  &  5h4m34s  &  15d24m14s  &  \cite{Auriere_v} \\

	HD40711  &  528$\pm$38  &  1245  &  6h1m1s  &  10d24m5s  & \cite{Auriere_v} \\

	HD90569  &  541$\pm$23  &  -  &  10h27m38s  &  9d45m44s  &  \cite{Auriere_v} \\

	HD94427  &  356$\pm$41  &  -  &  10h53m56s  &  -12d26m4s  & \cite{Auriere_v} \\

	HD103498  &  169$\pm$19  &  -  &  11h55m11s  &  46d28m11s  & \cite{Auriere_v} \\

	HD171586  &  375$\pm$56  &  -  &  18h35m36s  &  4d56m9s  & \cite{Auriere_v} \\

	HD171782  &  333$\pm$78  &  -  &  18h36m29s  &  5d17m19s  & \cite{Auriere_v} \\

	HD220825  &  312$\pm$25  &  -  &  23h26m55s  &  1d15m20s  & \cite{Auriere_v} \\

	HD66051  &  50$\pm$24  &  17.63011  &  8h1m24s  &  -12d47m35s  & \cite{Kochukhov2018} \\

	HD188774  &  49$\pm$15  &  -  &  19h55m10s  &  41d17m10s  & \cite{Neiner2015} \\

	HD151525  &  146$\pm$38  &  -  &  16h47m46s  &  5d14m48s  & \cite{Auriere_v} \\

	HD204411  &  88$\pm$14  &  -  &  21h26m51s  &  48d50m6s  & \cite{Auriere_v} \\

    \hline
 \end{tabular}
\end{table*}

\begin{table*}
\caption{List of O stars} \label{tab:O_stars}
\begin{tabular}{lcccccc}
    \hline
    Name & $B_{\mathrm{rms}}$ $\pm$ $\sigma_B$ & P$_{\mathrm{orbital}}$ & Ra & Dec & Ref. \\
    &  [G] & [days] & & &  \\
    \hline\\
    SMC 159-2  &  2780$\pm$990  &  -  &  0h53m29s  &  -72d41m44s  & \cite{Bagnulo17}  \\
    
    2dFS936  &  965$\pm$530  &  -  &  5h1m8s  &  -68d11m45s  &  \cite{Bagnulo17} \\

    HD47129  &  810$\pm$150  &  14.396257  &  6h37m24s  &  6d8m7s  & \cite{Grunhut13} \\

    HD54879  &  716.5$\pm$78  &  -  &  7h10m8s  &  -11d48m9s  &  \cite{Castro15} \\

    HD108  &  325$\pm$46  &  -  &  0h6m3s  &  63d40m46s  &  \cite{Shultz_17} \\

    NGC 1624-2  &  5920$\pm$2448  &  -  &  4h40m37s  &  50d27m41s  & \cite{Wade_2012} \\

    HD148937  &  210$\pm$71  &  8000  &  16h33m52s  &  -48d6m40s  & \cite{Wade2012}  \\

    CPD -28$^{\circ}$ 2561  &  392$\pm$274  &  -  &  7h55m52s  &  -28d37m46s  & \cite{Wade_15} \\

    HD37022  &  366$\pm$126  &  -  &  5h35m16s  &  -5d23m22s  &  \cite{Wade_2006} \\

    HD191612  &  486$\pm$137  &  1548.3  &  20h9m28s  &  35d44m1s  & \cite{Wade_11} \\

    HD57682  &  170$\pm$53  &  -  &  7h22m2s  &  -8d58m45s  & \cite{Grunhut09} \\

    \hline
 \end{tabular}
\end{table*}

\begin{table*}
\caption{List of weakly-magnetic (WM) stars} \label{tab:WM_stars}
\begin{tabular}{lccccccc}
    \hline
    Name & Evol. status & $B_{\mathrm{rms}}$ $\pm$ $\sigma_B$ & P$_{\mathrm{orbital}}$ & Ra & Dec & $\mathrm{v \sin{i} \pm \sigma_v}$ & Ref. \\
    & & [G] & [days] & & & [km/s]&  \\
    \hline\\
    HD5550  & Ap (MS) & 13$\pm$4  &  6.82054  &  0h58m31s  &  66d21m6s  &  4.7$^{+1.3}_{-2.8}$& \cite{alecian16} \\
    
    HD172167  & A0V (MS) & 0.6$\pm$0.2  &  -  &  18h36m56s  &  38d47m1s  & 24.5$\pm$1.4& \cite{Lignieres-2009} \\
    &&&&&&& \cite{Royer14} \\


    HD95418  & A1V (MS) & 1$\pm$0.8  &  -  &  11h1m50s  &  56d22m56s  & 46.2 $\pm$1.2 & \cite{Blazere16} \\
    &&&&&&& \cite{Royer14} \\

    HD97633  & A2V (MS) & 0.4$\pm$0.3  &  -  &  11h14m14s  &  15d25m46s  & 23$\pm$1 & \cite{Blazere16} \\
    &&&&&&& \cite{Royer2002} \\

    HD67523  & F5II & 0.29$\pm$0.32  &  -  &  8h7m32s  &  -24d18m15s  & 8$\pm$0.4 &  \cite{Neiner17}\\
    &&&&&&& \cite{AE2012} \\

    HD48915A  & A1V (MS) &  0.2$\pm$0.1  &  ?  &  6h45m8s  &  -16d42m58s  & 16$\pm$1& \cite{Petit11} \\
    &&&&&&& \cite{Royer2002} \\
    
    HD52089  & B2Iab & 7.3$\pm$5.9  &  -  &  6h58m37s  &  -28d58m19s  & 22$\pm$3 & \cite{Fossati15} \\
    &&&&&&& \cite{Fraser2010} \\
    
    HD219134  & K3V &  1.6$\pm$0.3  &  planets!  &  23h13m16s  &  57d10m6s  & - & \cite{Folsom2018} \\
    
    HD47105 & Am/Ap (MS, end)& 6.7$\pm$1.89 & 4614.51 &- &- & 12$\pm$1 & \cite{Blazere2020}\\
    \hline
 \end{tabular}
\end{table*}

\bsp	
\label{lastpage}
\end{document}